\documentclass[ aps, prb, reprint]{revtex4-1}
\usepackage{amsmath,amsbsy,amssymb,amsfonts}
\usepackage{graphicx, color}
\usepackage{empheq}
\usepackage{hyperref}
\usepackage[capitalise]{cleveref}
\usepackage{subfigure}

\usepackage{ulem}
\usepackage{soul}

\renewcommand{\vec}[1]{\pmb{#1}}

\begin{document}
\date{\today}

\title{Adiabatic monoparametric autonomous motors enabled by self-induced nonconservative forces}

\author{Arkady Kurnosov$^1$, Lucas J. Fern\'andez-Alc\'azar$^{2,*}$, Ra\'ul Bustos-Mar\'un$^{3,4}$, Tsampikos Kottos$^1$}
\affiliation{
$^1$Wave Transport in Complex Systems Lab, Department of Physics, Wesleyan University, Middletown, CT-06459, USA\\
$^{2}$Instituto de Modelado e Innovaci\'on Tecnol\'ogica (CONICET-UNNE) and Facultad de Ciencias Exactas, Naturales y Agrimensura,
Universidad Nacional del Nordeste, Avenida Libertad 5400, W3404AAS Corrientes, Argentina \\
$^3$Instituto de F\'isica Enrique Gaviola and Facultad de Matem\'atica, Astronom\'ia, F\'isica y Computaci\'on, Ciudad Universitaria s/n, 5000, C\'ordoba, Argentina\\
$^4$Facultad de Ciencias Qu\'imicas,  Universidad Nacional de C\'ordoba,  Argentina\\
$^*$ email: lfernandez@exa.unne.edu.ar}

\begin{abstract}
Archetypal motors produce work when two slowly varying degrees of freedom (DOF) move around a closed loop of finite area in 
the parameter space.  Here, instead, we propose a simple autonomous {\it monoparametric} optomechanical 
engine that utilizes nonlinearities to turn a constant energy current into a nonconservative mechanical force. The latter self-sustains 
the periodic motion of a mechanical DOF whose frequency is orders of magnitude smaller than the photonic DOF. We have identified 
conditions under which the maximum extracted mechanical power is invariant and show a new type of self-induced robustness
of the power production against imperfections and driving noise.
\end{abstract}
\maketitle

\section{Introduction}
The vital role of nano/micro-engines in the advancement of nanotechnology has been placing their development at the forefront of 
the recent research activity \cite{CKOF16,WSCGLL16,KZDHF99,TMJBKMKS11,
KRPMKHEF11,LCGJC20,PG13,CSGWSSBGPHM10,WNMMMRB12,GK14,RKSCPCP18,ZBM14,KBH21,KSMLR14,GFMLCD16,bustos2013,CCBE16}. 
Many of the reported achievements have been benefiting areas ranging from nanorobotics to molecular electronics,
\cite{CKOF16,WSCGLL16,KZDHF99,TMJBKMKS11,KRPMKHEF11,PG13,CSGWSSBGPHM10,WNMMMRB12}
and from spintronics to quantum measurements \cite{RS08,AvO15,NBLACBS06,STPBXPWBR10,BMC10}.  
Along these lines, the concept of current-induced forces for the realization of nano/microscopic adiabatic quantum motors 
has emerged within the framework of modern condensed matter physics \cite{DMT09,TDM10,BKVEO11,bustos2013,FBP15}.
At the most fundamental level, adiabatic quantum motors utilize the interference effects of the electron current going through them 
to produce useful work extracted from a mechanical degree of freedom (motor). The motor degrees of freedom (DOF) are 
assumed to be slow compared to the electronic DOF allowing for a mixed quantum-classical description of the resulting dynamics. 
The quantum-coherent nature of the fast electronic degrees of freedom and the associated interference effects induce an adiabatic 
reaction force to the slow mechanical DOF (MDOF) \cite{bustos2013,FBP15}. 
The work per cycle associated with such forces has geometric features,
characterized by the area encircled by the MDOFs in their parameters' space \cite{brouwer1998,PPSW18,BS20,BTTOFA20}. Consequently, when
there is just a single (non-rotational) classical DOF, these reaction forces are necessarily conservative,  
producing zero useful work at the end of an adiabatic cyclic variation of the MDOF. 

The implementation of nanomotors in the condensed matter framework requires that the nanomechanical device is connected to electron reservoirs
with a temperature or a voltage gradient among them that provide the transport current \cite{DMT09,TDM10,BKVEO11,bustos2013,FBP15,BC19}. 
Alternative driving schemes (e.g. ac driving \cite{SPTZ10,KXGF14,FYHFCZ03}) and energy sources 
include chemical energy \cite{CKOF16,WSCGLL16} or light \cite{KZDHF99,PG13,WNMMMRB12,FKK21}. In fact, the recent 
advancements in nanophotonics provide tantalizing opportunities for the realization of autonomous nanomotors that might 
surpass fundamental operational limitations. Specifically, new features and tools that are intrinsic to the photonics framework, 
like the presence of (self-induced) nonlinearities due to light-matter interactions or the possibility to engineer losses (or gain), etc., 
might turn out to be useful design elements for bypassing such constraints, like, the multiparametric 
nature of the MDOFs.

\begin{figure}
\hspace*{-0.25cm}  
\includegraphics[width =1.05\columnwidth ]{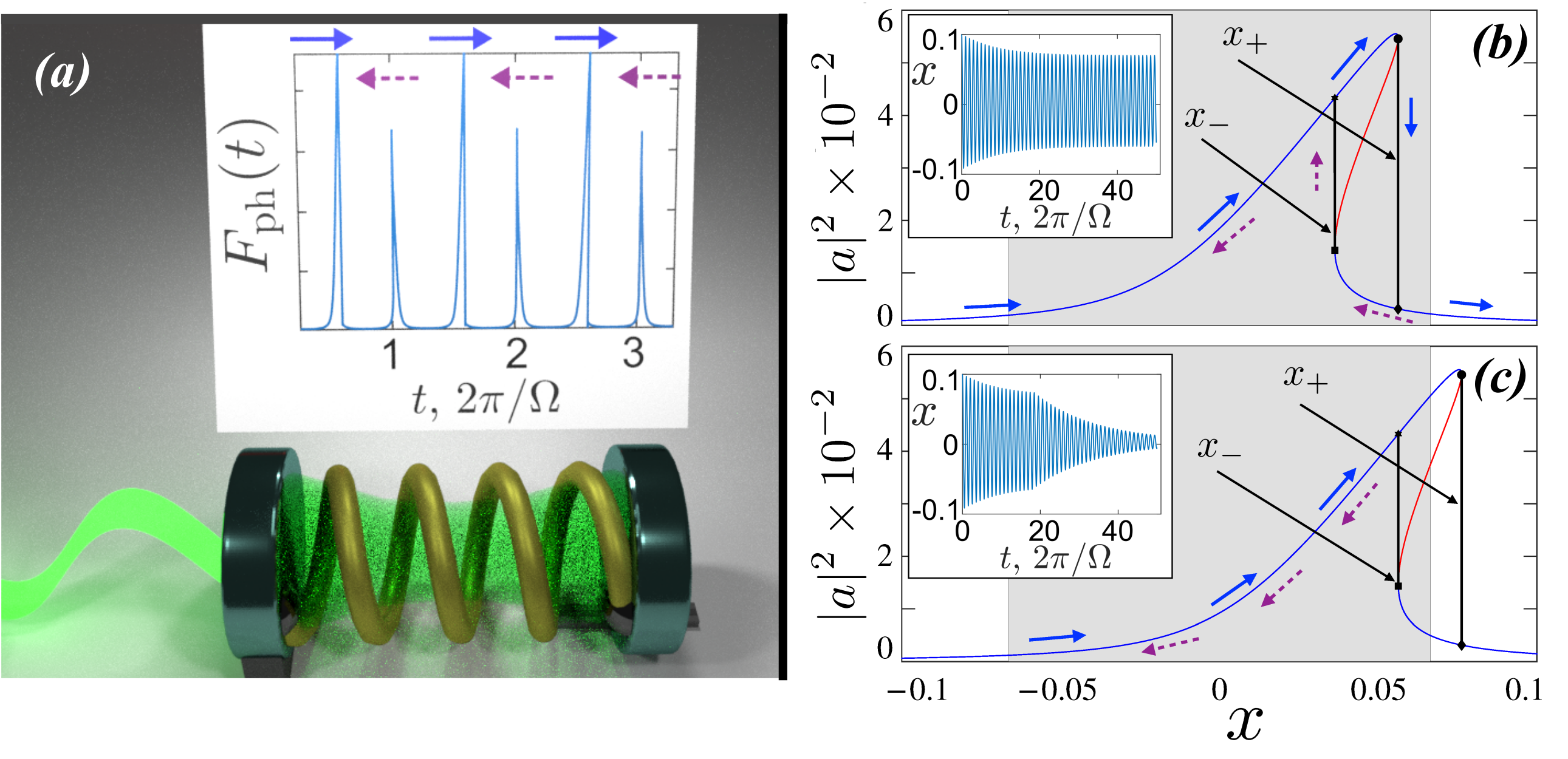}
\caption{\label{fig1}
$(a)$ Schematics of a single-cavity monoparametric motor. 
The radiation pressure in a nonlinear photonic cavity induces self-oscillations of the mirror.
({\bf Inset}) The magnitude of the photonic force $F_{\rm Ph.}\propto |a|^2$ depends on the mirror's direction of motion (see arrows)
due to the bistability of the modal energy $|a|^2$. 
$(b,c)$ The bistable dependence of the modal 
energy $|a|^2$ (blue lines) with the mirror's position $x$ [see Eq. (\ref{Eq:SingleCubic})] shows a hysteresis loop within the range 
$x\in(x_-,x_+)$. The red lines are real roots that do not correspond to physical solutions. The grey shaded area indicates
the range of oscillation of the MDOF. In $(b)$, the motion of the MDOF is wide enough to cover the whole loop and the modal energy $|a|^2$ 
explores both branches (see arrows). The work done on the MDOF is proportional to the area of the loop. In $(c)$, the 
amplitude of the MDOF cannot cover the loop and $|a|^2$ explores one branch of the bistability. 
[Insets (b,c)] Temporal dynamics of the MDOF that $(b)$  reaches the steady state; $(c)$ relaxes toward the equilibrium position.}  
\end{figure}

Here, we propose an autonomous {\it monoparametric} optomechanical motor consisting of a single harmonic MDOF coupled 
to a nonlinear photonic circuit driven by a monochromatic source. In the example shown in Fig. \ref{fig1}, the photonic 
circuit consists of a Fabry-P\'errot resonator, while the MDOF is described by an oscillating mirror attached to a spring. The 
position of the mirror controls the resonant frequency of the photonic DOF (PDOF) and subsequently the energy flux via the detuning 
from the monochromatic source. For incident power above a critical value,
the intrinsic nonlinearity of the cavity produces bistability in the PDOF and 
the access to one state or the other is determined by the characteristics of the MDOF's motion. Consequently, the optical force
is self-regulated by the position and direction of the MDOF and can become non-conservative, thus compensating for the 
mechanical friction and enforcing an oscillatory steady state of the mirror. 
We show that the optical force 
undergoes a self-induced hysteresis loop when the position of the mirror changes. The associated 
area of the loop gives the work done per cycle. 
Devices based on our mechanism 
are resilient against stochastic noise associated with the lasing source. We identify optimal designs and 
derive conditions under which the maximum extracted power is invariant under various design parameters. Our theory utilizes 
an adiabatic coupled mode framework, but its predictions are applicable for as long as there is a large time-scale separation 
between the mechanical and the PDOF. The theoretical results have been scrutinized against time-domain simulations 
with temporal coupled mode theory (CMT) models and realistic electromechanical platforms.

Importantly, nonlinearities may also arise in the context of quantum transport via the mean-field 
treatment of electron-electron interactions\cite{NEGRE2008},
which are essential in all density-functional-theory-based methods \cite{DFT64}.
Therefore, it is reasonable to envision the extrapolation of our results to the design of novel quantum devices.

\section{Basic principle using a single cavity setup}
We consider a nonlinear single-mode cavity driven by a monochromatic 
source. The dynamics of the electromagnetic field inside the cavity is described by a time-dependent CMT  
\begin{equation}\label{Eq:SinglePhotonic}
\dot{a} = i\left[\omega_{a}(X) + \chi|a|^{2}\right]a - \gamma a  + i \sqrt{2\gamma_e }s_{p} e^{i\omega t},
\end{equation}
where the field amplitude $a(t)$ is normalized such that $|a|^{2}$ is the energy inside the cavity and $|s_{p}|^{2}$ and 
$\omega$ represent the power of the incident wave and its frequency. Here, $\gamma_e$ is the decay rate toward the input-output channel, and $\gamma\ge \gamma_e$ is the total loss.
Finally, the term $\chi|a|^{2}$ describes a nonlinear frequency correction due to,
for instance, a Kerr effect. The modal resonant frequency $\omega_{a}(X)$ depends parametrically on the (slow) MDOF $X$. For 
concreteness, we adopt below a language associated with the Fabry-P\'erot example of Fig. \ref{fig1}. In this framework,
$X$ is the position of the mirror, $\omega_{a}(X) = \omega_{0} L/(X + L)\approx \omega_{0}(1 - X/L)$ where $\omega_0$ is 
the resonant frequency of the cavity in the equilibrium position of the mirror $X=0$. Below, we assume that $X\ll L$.
Nonlinearity could be provided by a filling medium that does not limit the MDOF dynamics, 
like a gas \cite{LHFL09} or a thin film \cite{film06}.

The MDOF is described by a dumped harmonic oscillator driven by a photonic force proportional to the energy inside the 
cavity, $F_{\rm ph} = \epsilon|a|^{2}$ \cite{aspelmeyer2014},
\begin{equation}\label{Eq:SingleMechanic}
M\ddot{X} = -2\Gamma M\dot{X} - KX + \epsilon|a|^{2}
\end{equation} 
where $M$, $K$, and $\Gamma$ are the mirror mass, the spring constant, and the friction coefficient. The coupling is 
$\epsilon=\zeta L^{-1}$, where the model dependent coefficient $\zeta$ takes the value $\zeta=1$ in case of Fabry-P\'erot resonators. 
It is convenient to rewrite Eqs. (\ref{Eq:SinglePhotonic},\ref{Eq:SingleMechanic}) in terms of 
$x = X/L$, $\Omega = \sqrt{K/M}$, and $\alpha = \epsilon/M$,
 \begin{subequations}\label{Eq:SingleCavity}
 \begin{empheq}[]{align}
&\dot{a} = i\left[\omega_{0}(1 - x) + \chi|a|^{2}\right]a - \gamma a  + i \sqrt{2\gamma_e}s_{p} e^{i\omega t},\label{Eq:S1}\\
&\ddot{x} = -2\Gamma \dot{x} - \Omega^{2}x + \alpha|a|^{2}.\label{Eq:S2}
\end{empheq}
\end{subequations}

\section{Self-induced nonconservative force}
We consider the situation where the MDOF is much slower than the PDOFs, i.e., 
$\Omega\ll \gamma\ll \omega_{0}$. In this adiabatic limit, $x$ in Eq. (\ref{Eq:S1}) can be treated as a parameter rather than a time-
dependent variable.
It is then possible to find an analytical solution to Eqs. (\ref{Eq:S1},\ref{Eq:S2}), by introducing the ansatz $a(t) = \tilde{a} e^{i
\omega t}$ \cite{SuppM}. Then, Eq. (\ref{Eq:S1}) reads 
\begin{equation}\label{Eq:SingleCubic}
z\left[(z - A)^{2} + B\right] - J = 0
\end{equation}
where  $z = \chi|a|^{2}/\gamma$, $A(x) = (\omega_{0}x + \delta\omega)/\gamma$, $B=1$, $J  = 2\chi|s_{p}|^{2}/\gamma^{2}$, 
and the frequency detuning of the emitting source is $\delta\omega = \omega - \omega_{0}$. 
Specifically, when the injected power $|s_{p}|^2$ exceeds a critical value $|s_p^{cr}|^2$
\begin{equation}
\label{s-crit}
|s_p|^2 > |s_{p}^{cr}|^2 = (4/3)^{3/2}\gamma^3/(2\gamma_e\chi), 
\end{equation}
Eq. (\ref{Eq:SingleCubic}) admits three solutions for $z\propto|a|^2$ in some displacement range $x\in(x_-,x_+)$. This bistable
behavior is associated with the formation of a self-induced ``hysteresis loop'' in the parameter space defined by the optical force
and the displacement, see Fig. \ref{fig1}$(b,c)$. 
We stress that both displacement bounds $x_{\pm}$ are byproducts of Eq. (\ref{Eq:SingleCubic}) and they do not 
depend on the parameters of the MDOF. Instead, they only depend on the parameters associated with the PDOF, the 
detuning $\delta\omega$, and the incident field amplitudes $s_p$.  \cite{SuppM}

Let us analyze in more detail the motion of the mirror. First, consider the case where the mirror is at $x < x_{-}$ and it  
moves expanding the cavity. The modal energy of the PDOF $|a|^2$ will change along the upper branch 
of the hysteresis loop until the critical position $x_{+}$ is reached, see Fig. \ref{fig1}b. At this point, the energy $|a|^2$ sharply 
decreases. In contrast, when the mirror starts at a position $x>x_{+}$ and moves to contract the cavity, $|a|^2$ 
changes following the lower branch of the hysteresis loop until it reaches the position $x_{-}$, where the modal energy $|a|^2$ sharply
increases, see Fig. \ref{fig1}b. In consequence, the optical force ${\vec F}_{\rm ph}\propto|a|^{2}{\hat X}$ has a 
magnitude that depends on the mirror's direction. The associated work performed by the radiation pressure, $W_{\rm ph} = \oint 
F_{\rm ph}d{X}=\iint_D\left(-\partial F/\partial|a|^2\right) dXd|a|^2$, results proportional to the area of the hysteresis loop. 
Notice that the nonlinearity introduces an additional effective DOF, $|a|^2$, that enables a nonzero area.
Since such an area is insensitive to the details of the parametric trajectory, 
providing a self-induced robustness of the work production.

Energy conservation implies that, in the steady state, the work of the optical force $W_{\rm ph}$ should balance 
the dissipated energy $W_{\rm fr}$ in each cycle.
Assuming $x(t) = x_{\rm eq} + x_{0}\sin\Omega t$ and $|x_{\rm eq}|\ll x_{0}$, we can estimate the mirror's amplitude $x_0$ from
$W_{\rm ph}=W_{\rm fr}$, where $W_{\rm fr} = 2\pi \Gamma\Omega M x_0^2 L^2$. Therefore $x_0 \approx \sqrt{\alpha w/(2\pi\Gamma\Omega)}$,
where $w$ is the area encircled by the hysteresis loop. 
This expression for $x_0$, however, becomes inconsistent whenever $x_{0} < |x_{\pm}|$,
because the extraction of useful work from the MDOF requires that the optical force $F_{\rm ph}$ explores both branches of the bistability. 
If the mirror reaches its rightmost position at $x_{\rm eq} + x_0<x_{+}$ and turns back,
the driving force remains on the same 
branch, as shown by the arrows in Fig. \ref{fig1}c; therefore $w=0$ and consequently $F_{\rm ph}$ becomes conservative.
Importantly, this is true even if the initial displacement exceeds $x_{+}$. 
In such a case, the motion of the mirror simply relaxes to the equilibrium position, $x_{\rm eq}$ (inset of Fig. \ref{fig1}c).
Thus, to guarantee the existence of the MDOF's stationary regime, one has to supplement the necessary condition \cref{s-crit} 
with a  sufficient condition 
\begin{equation}
x_{0} > \max(|x_{-}|, |x_{+}|), 
\label{MechCrit}
\end{equation} 
that ensures the exploration of the whole bistability region.

The optomechanical nanostructure of Fig. \ref{fig1}a constitutes an autonomous monoparametric motor that converts energy from 
a constant energy current, generated by coherent radiation, into mechanical work. The performance of such a motor can be quantified 
by the output power $P_{\rm out}$ and the efficiency $\eta$ which are
\begin{equation}
 P_{\rm out}=W_{\rm ph}\cdot\frac{\Omega}{2\pi}, \quad     \eta=\frac{P_{\rm out}}{P_{\rm in}}=\frac{W_{\rm ph}\cdot\frac{\Omega}{2\pi}}
{|s_p|^2} \propto \frac{\Omega}{\omega_0},
 \label{efficiency}
\end{equation}
where the input power $P_{\rm in}=|s_{p}|^{2}$. The efficiency, although small ($\Omega/\omega_0\ll1$), is comparable to that 
of other systems in the adiabatic limit \cite{FKK21}.

\begin{figure}
\center\includegraphics[width=0.5\textwidth]{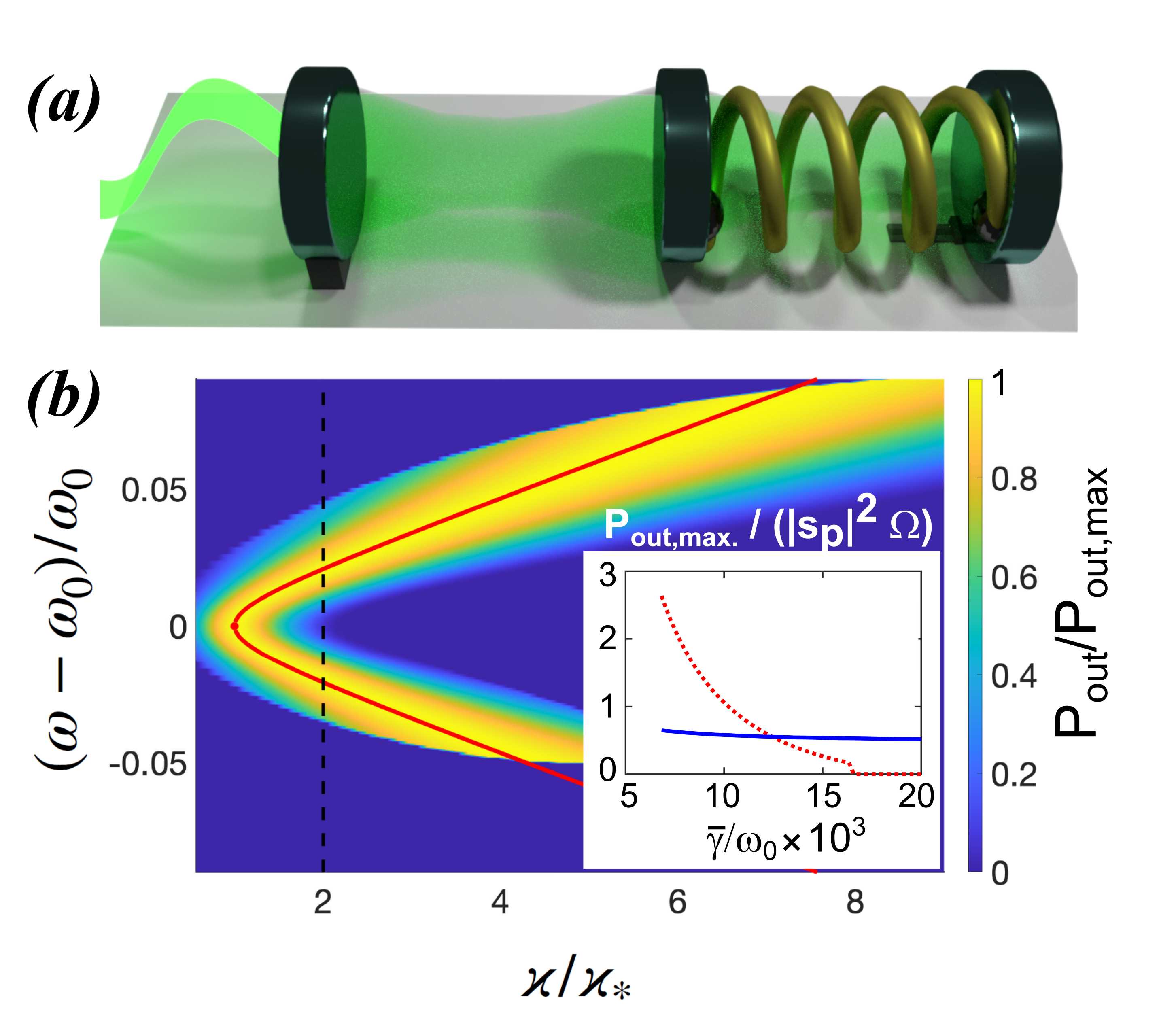}
\caption{ 
$(a)$ Schematics of the double-cavity monoparametric motor.
$(b)$ Output power versus emitter detuning $\delta\omega=\omega-\omega_0$ and coupling $\varkappa$, see Eq. (\ref{Eq:DoubleCavity}). 
The cutoffs are determined by Eq. (\ref{MechCrit}), and the solid red line indicates 
the maximum (iso)power line Eq. (\ref{Eq:Isopower}) with $\varkappa_{0} = \varkappa_{\ast}$. 
Vertical black dashed line indicates $\varkappa/\varkappa_*=2$ (see Fig. \ref{fig3}).
{\bf Inset:} Maximum power production of the single-cavity (red dotted line) and double cavity (blue solid line) motors vs.
the loss $\bar{\gamma}$.
The addition of a second cavity allows us to optimize the coupling $\varkappa$ to obtain  
a stable performance against losses.  (Parameters in Ref. \onlinecite{param1})
}
\label{fig2}
\end{figure}

The output power of the optomechanical nanomotor is affected dramatically by variations of the loss parameter $\gamma$. Our CMT analysis indicates that the performance of the motor 
deteriorates rapidly for moderate $\gamma$-values or even drops to zero due to the violation of Eq. (\ref{MechCrit}), 
see inset of Fig. \ref{fig2}$(b)$. Obviously, in any realistic scenario, such rapid performance deterioration  
is undesirable. It turns out that we can eliminate this deficiency by incorporating in our design an additional PDOF. The 
idea is to separate the PDOF that couples to the source from the PDOF that implements the non-conservative 
optical force. In this way, the first PDOF will act as a buffer, protecting the second PDOF (and consequently 
the induced $F_{\rm ph}$) from any variations occurring in $\gamma$.
We proceed with the analysis of such a structure.

\section{ Double cavity setup}
We consider two coupled optical modes, $a$ and $b$, where $b$ is driven by the monochromatic source, 
while $a$ is nonlinear and its resonant frequency depends on the position of the MDOF (see Fig. \ref{fig2}a). The associated 
equations of motion read
\begin{eqnarray}
 \dot{b} &=& i\omega_{0}b - \gamma_{b} b  + i\varkappa a + i \sqrt{2\gamma_{e}}s_{p} e^{i\omega t} \notag\\ 
 \dot{a} &=& i\left[\omega_{0}(1 - x) + \chi|a|^{2}\right]a - \gamma_{a} a  + i\varkappa b \label{Eq:DoubleCavity}\\
 \ddot{x} &=& -2\Gamma \dot{x} - \Omega^{2}x + \alpha|a|^{2},\notag
\end{eqnarray}
where $\varkappa$ is the coupling parameter between the modes and $\gamma_a,\gamma_b(\geq \gamma_e)$ are the decay rates of the modes. 

By applying the adiabatic approximation in Eqs. (\ref{Eq:DoubleCavity}), we 
arrive at a similar equation as Eq. (\ref{Eq:SingleCubic}) for the field intensity of the second PDOF. Now, $z = \chi|a|^{2}/
\bar{\gamma}$ where ${\bar \gamma}= (\gamma_{a} + \gamma_{b})/2$ and only $A$ depends on $x$ \cite{SuppM}.
Following the same steps as previously, we find the critical power $|s_p^{cr}|^2$ that ensures self-oscillations and analyze the 
hysteresis loop in the force-displacement plane to extract the equivalent mechanical criterion Eq. (\ref{MechCrit}) for work production. 

In this configuration, the coupling $\varkappa$ (together 
with the emitter detuning $\delta \omega$) controls the flow of energy to the second PDOF that forms the non-conservative 
force due to light-matter interactions. In Fig. \ref{fig2}$b$, we report the extracted power Eq. (\ref{efficiency}) for different 
values of $\varkappa$ and $\delta \omega$ keeping fixed the rest of the parameters. The 
red line indicates an ``isopower'' line where the output power remains the same. This curve is described by the equation 
\begin{equation}\label{Eq:Isopower}
(\omega - \omega_{0})^{2} =  (\gamma_{b}/\varkappa_{0})^{2}(\varkappa^{2} - \varkappa_{0}^{2})
\end{equation}
for any particular values of $\omega_0$, $\varkappa_{0}$, and $\gamma_b$ \cite{SuppM} and provides 
a desirable flexibility in designing the motor. 

To compare the performance of this setup with the one in Eq. \ref{Eq:SingleCavity}, we require three conditions: (a) The 
coupling to the source $\gamma_e$ in both setups is the same; (b) The average loss in each setup is the same, 
$\gamma={\bar \gamma}$; and (c) the source power $|s_{p}|^2$, characteristic frequency $\omega_{0}$, nonlinearity 
coefficient $\chi$, and all the mechanical parameters are the same. Under these conditions, we calculate the maximum value 
of $P_{\rm out}$ of the double cavity by exploring the $(\varkappa,\delta \omega)$-plane as a function of 
$\bar{\gamma}=(\gamma_a + \gamma_b)/2$, with $\gamma_a=cons.$ and $\gamma_e = \bar{\gamma}$.
For the single cavity, we adjust $\gamma_e = \gamma=\bar{\gamma}$. 
We find that the performance of 
the single-cavity setup (red dashed line) is better at high-quality factors ($\sim1/\bar{\gamma}$) while the double-cavity demonstrates 
a degree of robustness against variations in $\bar{\gamma}$ (see inset of Fig. \ref{fig2}b). 
Importantly, it does not vanish as $\bar{\gamma}$ increases. 

 \begin{figure}
\center\includegraphics[width=0.5\textwidth]{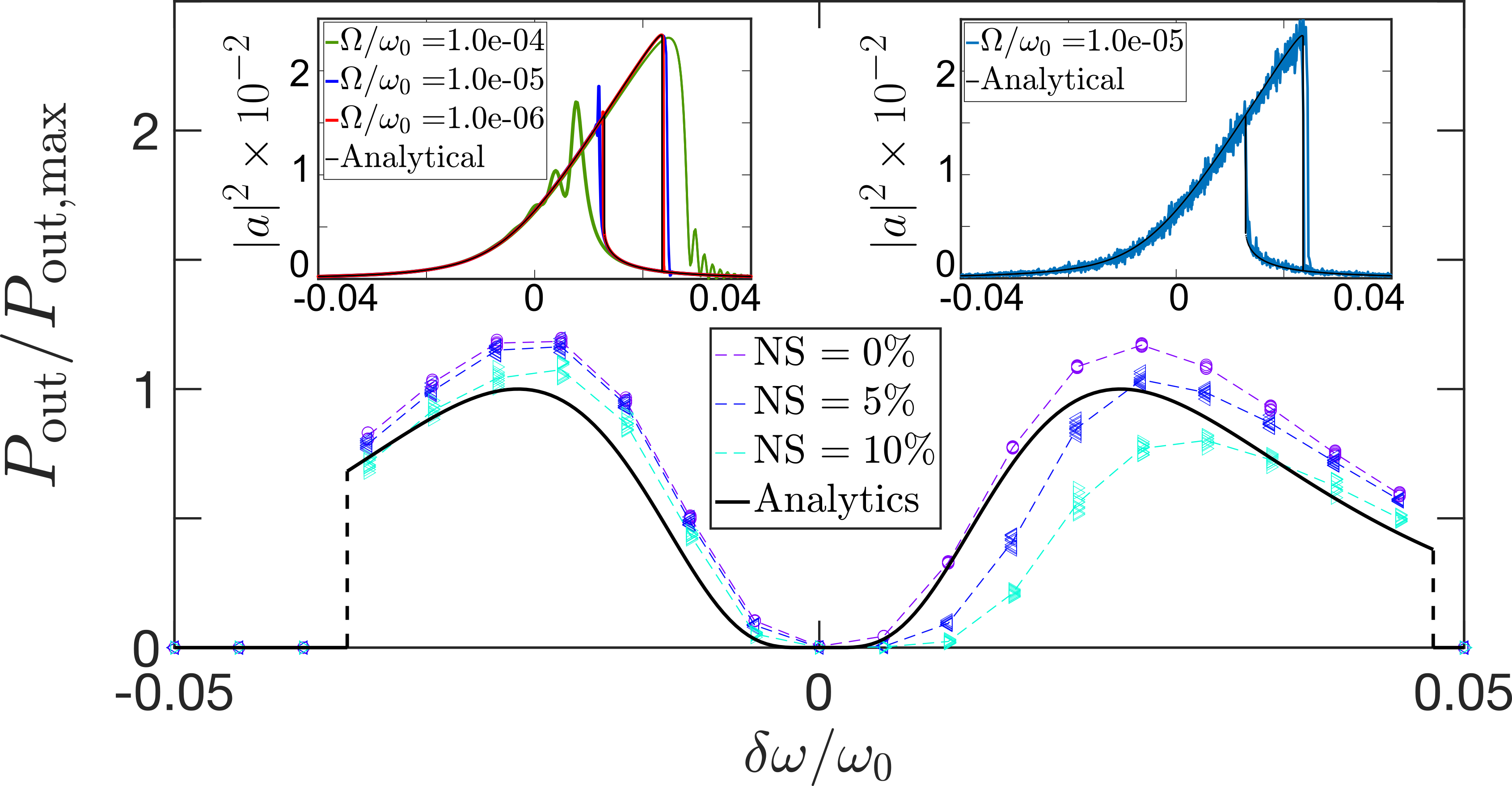}
\caption{
Output power of the double-cavity motor with coupling $\varkappa/\varkappa_*=2$ vs. $\delta \omega$ for different noise strengths $NS$.
Our time-domain simulations (with $\Omega/\omega_0=10^{-5}$) in the absence of noise (violet dashed line) deviate slightly from the adiabatic 
prediction (black solid line) because of dynamical effects. Strong noise strength $NS=5\%$ and $NS=10\%$ produce small deviations 
for $\delta\omega<0$. Symbols indicate values of different noise realizations.
(Left inset) The field intensity dynamics $|a(t)|^2$ vs $x(t)$ for $\Omega/\omega_0=10^{-4}$ (green line), 
$\Omega/\omega_0=10^{-5}$ (blue line), and
$\Omega/\omega_0=10^{-6}$ (red line) agree nicely with the analytical prediction (black line) 
(Right inset) Strong noise $NS=10\%$ does not destroy the hysteresis loop.
 Here $\Omega/\omega_0=10^{-5}$.
}
\label{fig3}
\end{figure}

\section{Dynamical simulations and noise} 
For practical implementations, it is necessary to assess the robustness of our proposal using realistic 
conditions, like the presence of noise or non-adiabatic (dynamical) effects. To this 
end, we consider time-domain simulations using the CMT modeling where the driven mode $b$ is affected by noise. 

In Fig. \ref{fig3}, we show the extracted power evaluated from dynamical simulations for a vertical cut of Fig. \ref{fig2}(b) 
with $\varkappa/\varkappa_*=2$ versus the detuning $\delta\omega/\omega_0$. First, we discuss the case of a finite value of
$\Omega/\omega_{0}=10^{-5}$ in the absence of noise ($NS=0\%$). We observe that the power from dynamical simulations is slightly 
higher than the analytical prediction. We attribute this small deviation to 
dynamical effects. In the left inset of Fig. \ref{fig3}, we show the field intensity $|a(t)|^{2}$
vs. $x(t)$ for different values of $\Omega/\omega_{0}$. The agreement with the adiabatic solution is satisfactory for  
$\Omega/\omega_{0}\lesssim 10^{-5}$ while a reasonable agreement persists for even higher ratios $\Omega/\omega_{0}=10^{-4}$. 

Next, we consider white noise associated with the driving source \cite{SuppM}.
The noise strength $NS$ is quantified by comparing the 
fluctuations of the optical field intensity, in terms of its variance $\sigma_b^2={\rm Var}(|b(t)|^2)$, relative to the mean modal energy 
$E_b=\langle |b(t)|^2 \rangle$, $NS:=\sigma_b/E_b$. In Fig. \ref{fig3}, we report the extracted power in case of relatively 
strong noise ($NS=5\%$ and $NS=10\%$). It turns out that even in the extreme case of $NS=10\%$,  the performance of the motor remains 
relatively unaffected. Such stability originates in that the hysteresis loop persists even 
when noise affects the photonic field (see right inset of Fig. \ref{fig3}).
%

\section{Electromechanical motor}

Here, we propose two designs of monoparametric electromechanical motors based on electronic circuit setups
that display a steady-state motion of a MDOF.
We validate our predictions via realistic time-domain simulations.

\begin{figure}
\hspace*{-0.35cm}  
\includegraphics[width=0.5\textwidth]{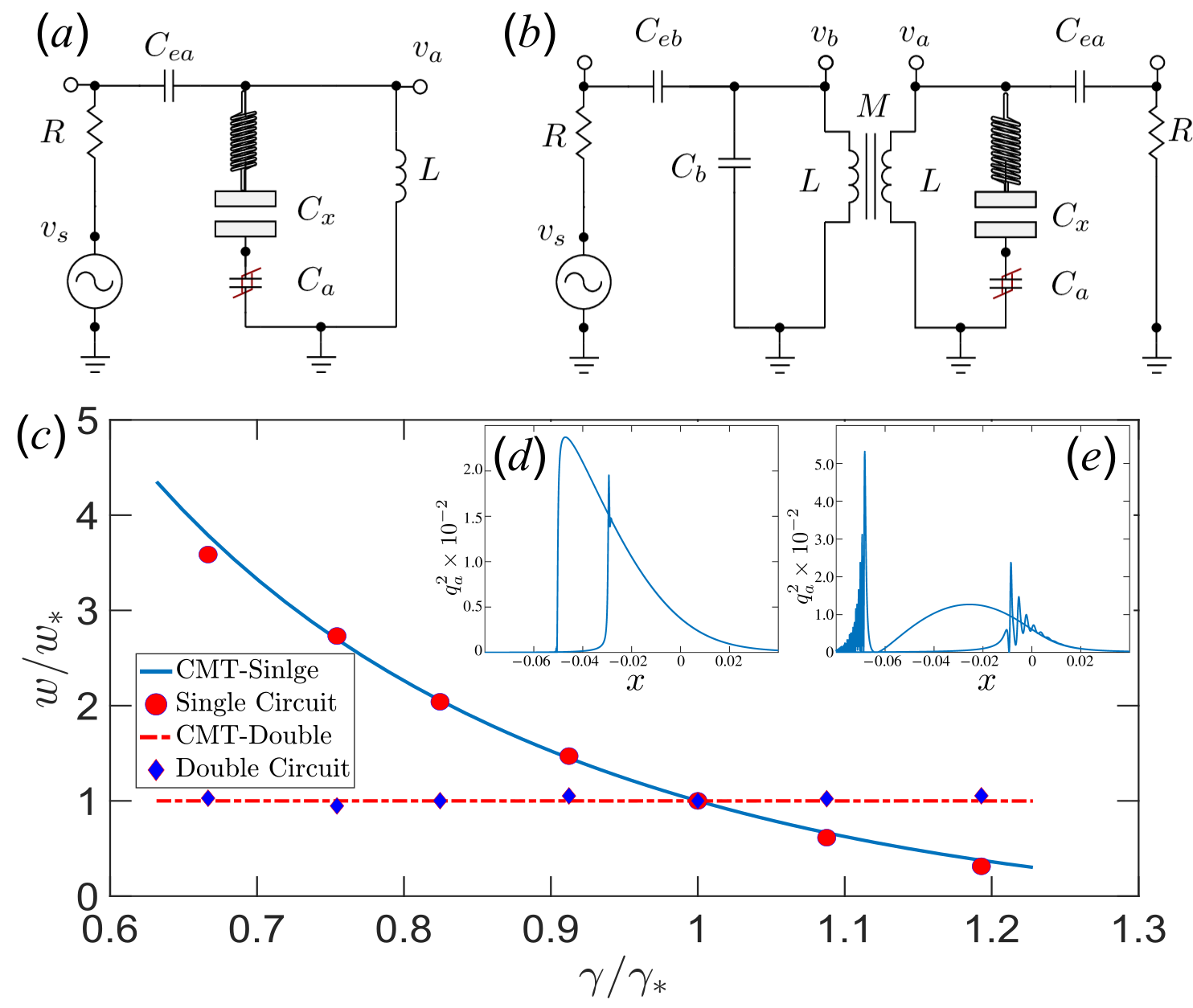}
\caption{\label{Fig4}Schematics  of (a) single-circuit and (b) double-circuit setups. 
The transmission line impedance is $R = 50\,\Omega$; the inductance $L$ and linear components of conductances 
$C_{a}$, $C_{b}$, $C_{x}$ satisfy $\sqrt{L/C_{0}} = z_{0} = 70\,\Omega$. $v_{s}(t) = v_{0}\sin\omega t$.
$(c)$ Normalized work production as a function of normalized electric field energy loss for single-circuit (red circles),
double-circuit (blue diamonds) and corresponding CMT approximations (blue solid and red dot-dashed lines respectively).
The reference loss value  chosen is $\gamma_{\ast} = 5.7\cdot 10^{-3}$, while the work 
$w(\gamma)$ is computed independently for each setup. Here, $w_{\ast}=w(\gamma_{\ast})$. One can see the single-circuit (single-cavity) 
setup performance is very sensitive to changes in $\gamma$ while double-circuit (double-cavity) setup is stable.
$(d,e)$ Charge squared $|q_a|^2$ vs. displacement $x$ during a whole cycle of the MDOF that evidences the hysteresis loop
associated to the single resonator $(d)$ and double resonator $(e)$ circuit. 
}
\end{figure}

As shown schematically in Fig. \ref{Fig4}$(a,b)$, we consider the electric circuit analogue of the single- and double- cavity setups 
consisting in one LC resonator and two coupled LC resonators, respectively.
In both setups, the nonlinearity is introduced by a nonlinear capacitor $C_{a}$ 
whose inverse capacitance depends nonlinearly on the charge $q$ as $C_{a}^{-1}(q) = C_{a0}^{-1} + \beta q^{2}$ 
with $C_{a0},\beta={\rm cons.}$ \cite{G99}. The circuit element $C_{x}$ is a parallel plate capacitor with one movable massive plate 
attached to a spring. Its associated capacitance depends on the plate displacement $\delta$ as
$C_{x}^{-1} = (d_{0} + \delta)/(\epsilon_{0} A) = C_{x0}^{-1}(1 + x)$, where $A$ is plate area, $d_{0}$ is the capacitor width 
in absence of bias and $x = \delta/d_{0}$ is a dimensionless displacement, considered as the MDOF. Therefore, the voltage on the node $v_{a}$ is
$v_{a}(q) = \frac{q}{C_{0}}(1 + \xi x) + \beta q^{3}$, where $C_{0}^{-1} = C_{x0}^{-1} + C_{a0}^{-1}$, $\xi = C_{0}/C_{x0}$.
Each LC resonator supports a resonant mode that, in absence of nonlinearity and displacement of the mechanical plate, 
has a resonance frequency $\omega_{0} = 1/\sqrt{LC_{0}} = 2\pi\times 10^9 rad/s$ and impedance at resonance $z_{0} = \sqrt{L/C_{0}}=70\,\Omega$.\cite{FKLK21}

The signal $v_s(t)=v_0\sin(\tilde{\omega} t)$ is generated by a voltage source connected to the circuit via a transmission line (TL) with 
characteristic impedance $R = 50\,\Omega$ ended in a coupling capacitance $C_{e\alpha}= \varepsilon_{\alpha}C_{0}$, $\alpha=a,b$. 
In contrast to the single cavity setup, where the TL is coupled to the  nonlinear mechanical LC resonator, in the 
double cavity setup the TL is coupled to a linear LC resonator via a capacitance $C_{eb}$ and, in turn, such resonator is coupled 
to the nonlinear mechanical LC resonator via a mutual inductance coupling with coefficient $\mu$. The later is also coupled to a TL 
that introduces an energy leakage controlled by the coupling capacitor $C_{ea}$.
The capacitive coupling to the TLs introduces a frequency shift. To avoid impedance mismatch between the resonators in the double cavity setup,
we keep their resonant frequencies approximately  the same by introducing the conditions $C_{b} = (1 - \lambda)C_{0}$ and 
$\varepsilon_{b} - \lambda \simeq \varepsilon_{a}$.

The back action to the MDOF is determined by the force induced by the electric field $E$ 
inside the capacitor $C_{x}$ on the movable plate's charge, $F = -E q = -[q/(\epsilon_{0}A)]q = -q^{2}\xi/(d_{0}C_{0})$. 
Such a force displaces the capacitor plate according to the dimensionless equation 
\begin{equation}
\frac{d^{2}{x}}{d\tau^{2}} = - 2{\Gamma}\frac{dx}{d\tau} - {\Omega}^{2}{x}  - {\alpha}q_{a}^{2} ,
\end{equation}
where $\tau=t/\omega_0$, $\Omega = \omega_{0}^{-1}\sqrt{K/M} \simeq 10^{-5}$, $\alpha = \xi v_{0}^{2}/(Md_{0}^{2}\omega_{0}^{3})$, and 
$q_a=q/(C_0 v_0)$.

The coupling of the resonators with TLs produces an electric energy leakage with decay rates $\gamma_{b}$ and $\gamma_{a}$ when $\mu = 0$.
Such rates can be determined, e.g. by estimating transient decay times, or alternatively finding the linewidths in a scattering analysis.
In addition, those decay rates can be approximated via a CMT description of the circuit setups, as described in
Ref. \onlinecite{FKLK21}. There, the CMT predicts decay rates $\gamma_{a, b}\propto\varepsilon_{a, b}^{2}(z_{0}/R)$ 
and therefore they can be controlled by varying the capacitive coupling to the TL.

In Fig. \ref{Fig4}$(c)$ we plot the (dimensionless) work production $w= \int|q_{a}|^{2}dx$ versus the decay rate $\gamma$, 
which is $\gamma=\gamma_a$ for the single resonator and $\gamma=\gamma_b$ for the double resonator setup. Variations of these
parameters are achieved by sweeping $\varepsilon_{a}$ and $\varepsilon_b$ in the interval $[0.085, 0.11]$, respectively.
We have normalized both the work production and the loss parameter $\gamma$ by a specific --but arbitrary-- set of parameters.
Specifically, for the double-resonator circuit 
we set $\varepsilon_{a} = 0.03$, $\varepsilon = 0.1$, which corresponds to losses 
$\gamma_{a} = 0.5\cdot 10^{-3}$, $\gamma_{b} = \gamma_{\ast} = 5.7\cdot 10^{-3}$.
We have found the maximum work $w$ by varying the coupling, $\mu$, and the driving frequency, ${\tilde \omega}$.
For the single-resonator circuit we choose resonant driving.
This approach allows us to compare $w/w_{\ast}$ vs. $\gamma/\gamma_{\ast}$, where $w_*=w(\gamma_*)$, for both circuit setups and their associated CMT.
One can see that the CMT accurately predicts the outcome of the time domain simulations of the electromechanical equations
also evidencing the robustness of the double-cavity performance while the single-cavity setup is sensitive to the parameters.

For completeness, in Figs. \ref{Fig4}$(d,e)$ we show the hysteresis loops associated to the charge square $|q_{a}|^{2}$
when $x$ changes for the single-resonator \ref{Fig4}$(d)$ and double-resonator \ref{Fig4}$(e)$ circuits.
Qualitative similarities with the photonic counterparts can be found for the single circuit setup, Fig. \ref{Fig4}$(d)$, although it is
not the case for the double circuit setup, \ref{Fig4}$(e)$. Despite this, in both cases, the loops enclose nonzero areas
and the results of circuit simulations shows an excellent matching with the CMT.

\section{Comparison with existing self-oscillations in optomechanical systems}

Self-sustained oscillations of the MDOF in single-cavity optomechanical systems have been known for some time  
and reported in a variety of theoretical and experimental platforms (see Ref. \cite{aspelmeyer2014} and references therein).
Here, we compare our proposal with the standard self-oscillations in optomechanical systems.
In principle, such setups can be described by the same equations of motion Eqs. (\ref{Eq:SingleCavity}) utilized here,
with the crucial difference that these setups do not involve a Kerr nonlinear correction to the frequency, i.e., $\chi\equiv 0$.  
We discuss next how self oscillations can develop.

We stress again that the optical force provides the energy required to sustain the motion of the MDOF,
and such energy can be interpreted as the area below the $(x,|a|^2)$-trajectory, i.e., $W_{\rm ph}\propto\oint |a|^2 dX$.
It turns out that in the adiabatic limit, $\Omega \ll \gamma\ll \omega_0$, and for $\Omega\to 0$, the field reacts 
instantaneously to the motion of the MDOF, and so does the force. Under this condition, the energy of the cavity
only depends on the position of the MDOF, $\left|a\right|^{2}\equiv\left|a\left(x\right)\right|^{2}$, which
implies that no net work can be produced when considering a closed trajectory, i.e. $W_{\mathrm{ph}}=0$.

One possibility to overcome this limitation corresponds to operating the device under {\it non-adiabatic} effects, which has been 
extensively studied both experimentally and theoretically \cite{Carmon05,rokhsari2005,kippenberg2005,Marquardt2006,zaitsev2011,aspelmeyer2014,guha2020}.
This is indeed a mechanism exploited in optomechanical platforms to create self-oscillations, 
which are enabled by the (extremely) high experimental quality factors , i.e., $\gamma, \Gamma \ll \Omega$. 
In these scenarios, the time that photons spend inside the cavity is enough to undergo the effect of the motion of the MDOF. 
As a result, the associated energy inside the cavity $\left|a\right|^{2}\equiv\left|a\left(x,t\right)\right|^{2}$ 
will draw a finite area in the $(x,|a|^2)$ plane, and, for appropriate input frequencies, a self-oscillation can be achieved.
This mechanism can also lead to more complex scenarios, such as dynamical multistability
(caused by the nonlinear optomechanical coupling and the high quality factor), which, in the case of high input power, 
results in chaotic MDOF dynamics.\cite{Carmon05} 
Of course, many realistic platforms cannot benefit from the high quality factors that are required to exploit the 
non-adiabatic effects, which limits the range of applications to a few experimental setups.

An alternative that can induce the MDOF motion consists of the introduction of another degree of freedom which can cause 
retardation effects between the photonic force $F_{\rm ph}\propto|a(X,t)|^2$ and the MODF $X$. 
One of such cases corresponds to the retardation produced by thermal effects where the heat has to propagate 
through a cantilever before it bends \cite{Hohenberg04}. 

In the present work, our approach is different from the ones discussed above.
While our mechanism focuses on the adiabatic limit, useful for a variety of platforms with moderate or even low 
quality factors, it does not consider retardation effects, which makes analytical calculations straightforward.
The idea is that, due to the additional (Kerr-type) nonlinear effects, the photonic force $F_{\rm ph}\propto|a|^2$
can distinguish the direction of motion of the MDOF, and  can be treated
as a function of position $x$ and direction of motion ${\dot x}/|{\dot x}|$ rather than a function of position only.
Such a dependence originates in the nonlinearity affecting the energy of the cavity that, under certain conditions, 
can show bistabilities leading to the formation of a hysteresis loop, as shown in Fig. \ref{fig1}.
A crucial feature of our proposal corresponds to the magnitude of the output power that can be delivered to the MDOF
that goes as $\Omega$, as opposed to non-adiabatic optomechanical self-oscillations that can deliver a power 
that is proportional to higher orders in $\Omega$ (larger than linear)
(see the \texttt{Supplementary material}\cite{SuppM}). 

Our proposal also differs from existing examples of self-oscillators in the framework of mechanics,
which are typically describing scenarios where the adiabatic approximation is not applicable
\cite{jenkins2013,serragarcia2018}.

\section{Conclusions}
We have introduced a novel class of adiabatic autonomous nonlinear motors that produce useful work based on the motion 
of a single MDOF with a clear separation of time scales from the PDOF. The monoparametric nature of these motors challenges
the common belief that the extraction of a useful work requires the control of (at least) two independent parameters; 
instead, the nonlinearity enables an effective extra DOF. The proposed motor offers: (a) A novel type of self-induced robustness that 
guarantees a work production that is resilient against driving noise and imperfections; (b) An opportunity to invoke MDOF and PDOF,
which do not have comparable frequencies. Mechanical modes with much lower frequency can now be utilized with our mechanism; and 
(c) The implied adiabaticity offers additional robustness against noise due to a self-averaging process of the trajectory of the MDOF.
We also validated our proposal via realistic time-domain simulations with an electromechanical version of the proposed
motor. It will be interesting to extend the concept of adiabatic autonomous monoparametric motors to solid-state systems\cite{NGS08}. 
A promising platform is ferromagnetic films, where nonlinearities due to the coupling of a macroscopic magnetic moment with lattice 
phonons can naturally emerge\cite{BALNO12}.

\begin{acknowledgments}
AK and TK acknowledge financial support by NSF-CMMI grant No. 1925543 and to Simons Foundation for Collaboration in MPS No. 733698.
LJFA acknowledges support by CONICET grant No. PIP2021: 11220200100170CO, UNNE grant No. PI: 20T001.
RBM acknowledges CONICET, SECYT-UNC, and ANPCyT grant No. PICT-2018-03587.
\end{acknowledgments}

\newpage

\onecolumngrid 

\begin{center}
\textbf{\large Supplementary Material}
\end{center}

\setcounter{equation}{0}
\setcounter{figure}{0}
\setcounter{table}{0}
\setcounter{page}{1}
\setcounter{section}{0}
\makeatletter
\renewcommand{\theequation}{S\arabic{equation}}
\renewcommand{\thefigure}{S\arabic{figure}}
\renewcommand{\thepage}{S\arabic{page}}
\renewcommand{\bibnumfmt}[1]{[S#1]}
\renewcommand{\citenumfont}[1]{S#1}


\section{\label{Sec:Reduction}Reduction of the equations of motion to the algebraic cubic equation}
In this section, we discuss the steady state solutions of the photonic degrees of freedom (PDOF) 
when the mechanical degree of freedom (MDOF) is considered as a parameter. We show that the 
photonic equations of motion reduce to a cubic equation whose solutions indicate the presence of bistability.

\subsection{\label{Sec:SC}Single-cavity}
We start by considering the time-dependent equations of motion 
 \begin{subequations}\label{Eq:SingleCavityRepeat}
 \begin{empheq}[]{align}
&\dot{a} = i\left[\omega_{0}(1 - x) + \chi|a|^{2}\right]a - \gamma a  + i s_{p}\sqrt{2\gamma_{e}} e^{i\omega t}\label{Eq:SC1}\\
&\ddot{x} = -2\Gamma \dot{x} - \Omega^{2}x + \alpha|a|^{2}.
\label{Eq:SC2}
\end{empheq}
\end{subequations} 
Here, the nonlinear optical cavity mode $a$ has a natural frequency $\omega_{0}$, is driven by a monochromatic source 
with frequency $\omega$ and source power $|s_{p}|^{2}$, and is coupled to a MDOF $x$, which is dimensionless. 
The field amplitude $a(t)$ is normalized such that $|a(t)|^{2}$ is the energy stored in cavity ``a'', 
$\gamma\ge \gamma_e$ is the loss, while $\chi|a|^{2}$ is a nonlinear 
frequency correction (we consider only the case of $\real{\chi}$). Here, $\Omega\ll\gamma\ll\omega\sim\omega_0$ and $\Gamma$ are MDOF frequency and loss,
while $\alpha$ is the reduced nonlinear coupling between the PDOF and the MDOF. 
 
The condition $\Omega\ll\gamma$ implies that the MDOF is very slow as compared with 
the electromagnetic field dynamics and justifies an adiabatic approximation, namely, treating \cref{Eq:SC1} as if $x$ were a constant parameter. 
To reformulate this condition, one can split slow and fast dynamics by the ansatz $a(t) = \psi(t)e^{i\omega t}$ ($|\psi|^{2} = |a|^{2}$), 
where the slow component $\psi(t)$ parametrically reflects the dynamics of $x(t)$. In the equation
\[
\dot{\psi} + i(\delta\omega + \omega_{0}x - \chi|\psi|^{2})\psi + \gamma \psi = i s_{p} \sqrt{2\gamma_{e}},
\]
where we introduce the laser detuning $\delta\omega = \omega_{0} - \omega$, one can omit the small term $\dot{\psi}\propto\Omega$.
By introducing $z = \chi|a|^{2}/\gamma$,  we obtain a cubic equation
\begin{equation}\label{Eq:SCalgebr}
\left[\left(z - A\right)^{2} + B\right]z - J = 0,
\end{equation}
with parameters
\begin{equation}\label{Eq:Params}
\begin{split}
&A(x) = (\omega_{0}x + \delta\omega)/\gamma,\\
&B\equiv 1,\\
&J = 2\gamma_{e}\chi|s_{p}|^{2}/\gamma^{3}.
\end{split}
\end{equation}
The time dependence in the latter equation becomes implicit. From its analysis in \cref{Sec:Z3solve}, we will explore the parametric 
dependence of the modal energy $|a|^{2}$ stored in the cavity as a function of $x$. 

\subsection{\label{Sec:DC}Double-cavity}
\begin{figure}
\center\includegraphics[scale = .6]{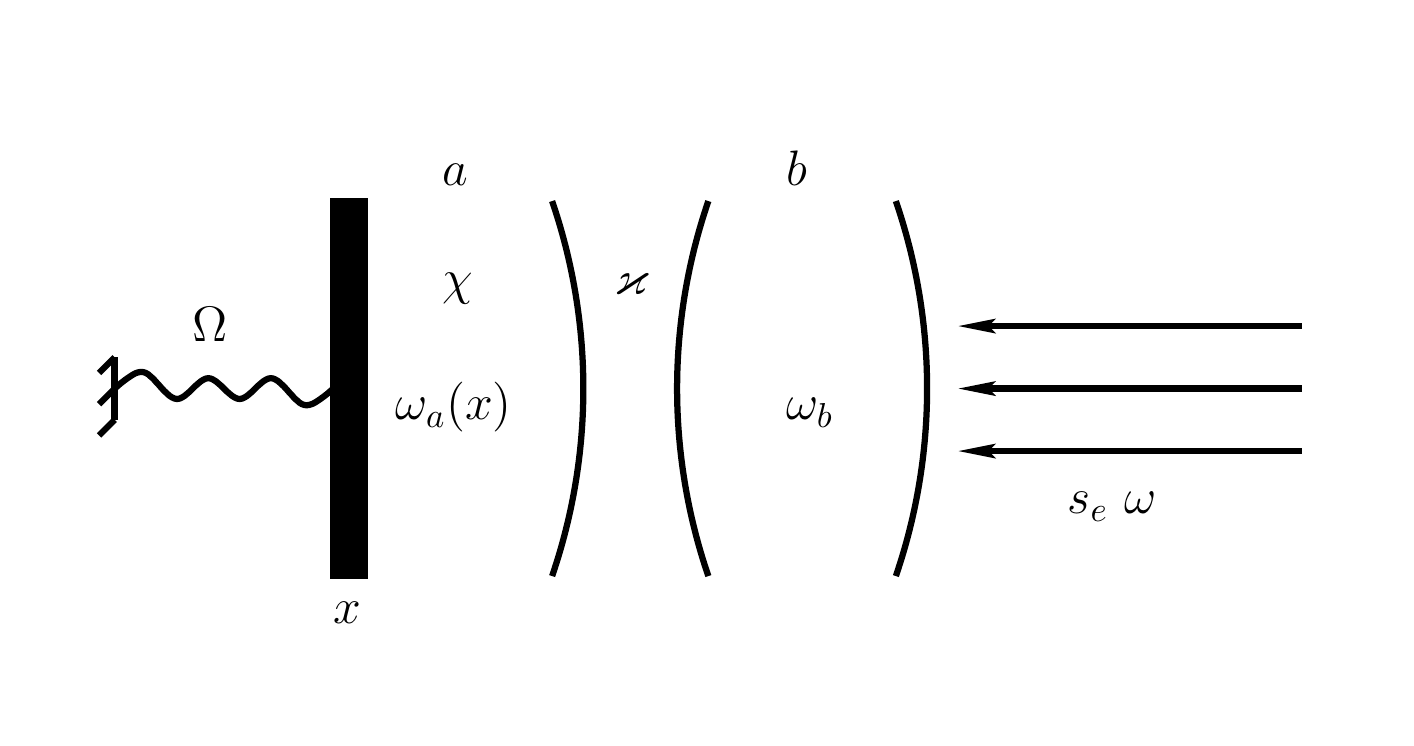}
\caption{\label{Fig:SetUpMonochrom}Schematic representation of the double cavity setup.}  
\end{figure}
Next, we consider the system depicted in \cref{Fig:SetUpMonochrom}. There, a monochromatic laser with frequency $\omega$
is directed toward the photonic cavity $b$, with natural frequency $\omega_b=\omega_{0}$, that is coupled to the nonlinear photonic mode $a$. 
The latter is coupled to a movable mirror attached to a spring. The natural frequency of the cavity mode $\omega_{a}$ is $\omega_{0}$
at low field intensities and when the position of the left mirror is fixed at $x = 0$. 
Using coupled modes theory (CMT) for the PDOF, we describe the system via the equations
\begin{subequations}\label{Eq:SetUp}
 \begin{empheq}[left=\empheqlbrace]{align}
&\dot{b}  = i\omega_{0}b - \gamma_{b}b + i\varkappa a + i \sqrt{2\gamma_{e}}s_{\rm p}e^{i\omega t}\label{Eq:1}\\ 
&\dot{a} = i\omega_{a}(x)a - \gamma_{a}a + i\varkappa b + i\chi|a|^{2}a\label{Eq:2}\\
&\ddot{x} = -2\Gamma \dot{x} - \Omega^{2}x + \alpha|a|^{2},\label{Eq:3}
 \end{empheq}
\end{subequations}
were $\gamma_{a}$ and $\gamma_{b}-\gamma_e$ are the losses due to radiation in modes $a$ and $b$, 
$\gamma_e$ is the loss due to the coupling of mode $b$ to the continuum, and $\Gamma$ is the mechanical friction. 
The emitter amplitude $s_{p}$ is such that $|s_{p}|^{2}$ is the incident source power, while $|a|^{2}$, $|b|^{2}$ is 
the energy stored in the respective mode. 

The frequency of the mode $a$ is inversely proportional to the length of the cavity and then, it is modulated
by the mechanical degree of freedom (MDOF) $x$, which represents a small displacement of the mirror normalized 
by the cavity length. 
Such a frequency reads
\begin{equation}
\omega_{a}(x) = \frac{\omega_{0}}{(1 + x)}\approx \omega_{0}(1 - x), \quad x \ll 1.
\end{equation}
This linear frequency approximation is also useful to describe frequency modulations by MDOFs in other platforms of interest,
like the electronic circuits discussed later on.

Using the condition $\Omega/\omega_{0}\ll 1$, we use the adiabatic approximation, as in \cref{Sec:SC}, 
treating $x$ in \cref{Eq:1,Eq:2} as a parameter. Then, we solve the first two equations in \cref{Eq:SetUp} 
 \begin{subequations}\label{Eq:SetUpRevisit}
 \begin{empheq}[left=\empheqlbrace]{align}
&\dot{b}  = i\omega_{0}b - \gamma_{b}b + i\varkappa a + i s_{\rm e}e^{i\omega t}\label{Eq:r1}\\ 
&\dot{a} = i\left[\omega_{0}(1 - x) + \chi|a|^{2}\right]a - \gamma_{a}a + i\varkappa b,
\label{Eq:r2}
 \end{empheq}
\end{subequations}
considering a parametric modulation via $x$. Here, $s_e=\sqrt{2\gamma_e}s_p$.

We look for stationary solutions of \cref{Eq:SetUpRevisit} of the form $(a, b)=(a_0,b_0) \exp\left\{i\omega t\right\}$.
The equation could be written in matrix form as   
\begin{equation}\label{Eq:FreqDomain}
\begin{bmatrix}
i\delta\omega +  \gamma_{b} & - i\varkappa\\
- i\varkappa &i(\omega_{0}x - \chi|a_0|^{2} + \delta\omega) + \gamma_{a}
\end{bmatrix}
\begin{bmatrix}
b_0\\
a_0
\end{bmatrix}
=
\begin{bmatrix}
i s_{e}\\
0
\end{bmatrix},
\end{equation}
where the emitter detuning $\delta\omega = \omega - \omega_{0}$. It follows that 
\begin{equation}\label{Eq:abs2}
|a_0|^{2} = \frac{\varkappa^{2}|s_{\rm e}|^{2}}{\Gamma_{b}^{2}\left[\omega_{0}x - \chi|a_0|^{2} - \xi_{\omega}\right]^{2} + \eta_{\omega}^{2}},
\end{equation}
where we introduced auxiliary notations
\begin{equation}\label{Eq:aux}
\begin{split}
&\Gamma_{b}^{2} = \delta\omega^{2} + \gamma_{b}^{2},\\
&\xi_{\omega} = -\delta\omega\left[1 - \frac{\varkappa^{2}}{\delta\omega^{2} + \gamma_{b}^{2}}\right],\\
&\eta_{\omega}^{2} =  \left(\varkappa^{2} + \gamma_{a}\gamma_{b}\right)^{2} + \left[\gamma_{a}^{2} - \frac{\varkappa^{4}}{\Gamma_{b}^{2}}\right]\delta\omega^{2},\\
&\bar{\gamma} = \frac{\gamma_{a} + \gamma_{b}}{2}.
\end{split}
\end{equation}
By introducing  the variable $z = \chi|a_0|^{2}/\bar{\gamma}$, we may rewrite \cref{Eq:abs2} in the cubic form
\begin{equation}\label{Eq:GofZ}
G(z) = z\left[(z - A)^{2} + B\right] - J = 0,
\end{equation}
where now the coefficients
\begin{equation}
\begin{split}\label{Eq:DCParams}
&A(x) = \frac{\omega_{0}x - \xi_{\omega}}{\bar{\gamma}},\\
&B = \frac{\eta_{\omega}^{2}}{\bar{\gamma}^{2}\Gamma_{b}^{2}},\\
&J = \frac{\varkappa^{2}|s_{e}|^{2}\chi}{\bar{\gamma}^{3}\Gamma_{b}^{2}}.
\end{split}
\end{equation}
Because of formal equivalence between \cref{Eq:SCalgebr} and \cref{Eq:GofZ}, the analysis of the next section where we discuss their solutions, is applicable to both cases. 

\subsection{Algebraic cubic equation\label{Sec:Z3solve}}
\begin{figure}
\center\includegraphics[scale = .18]{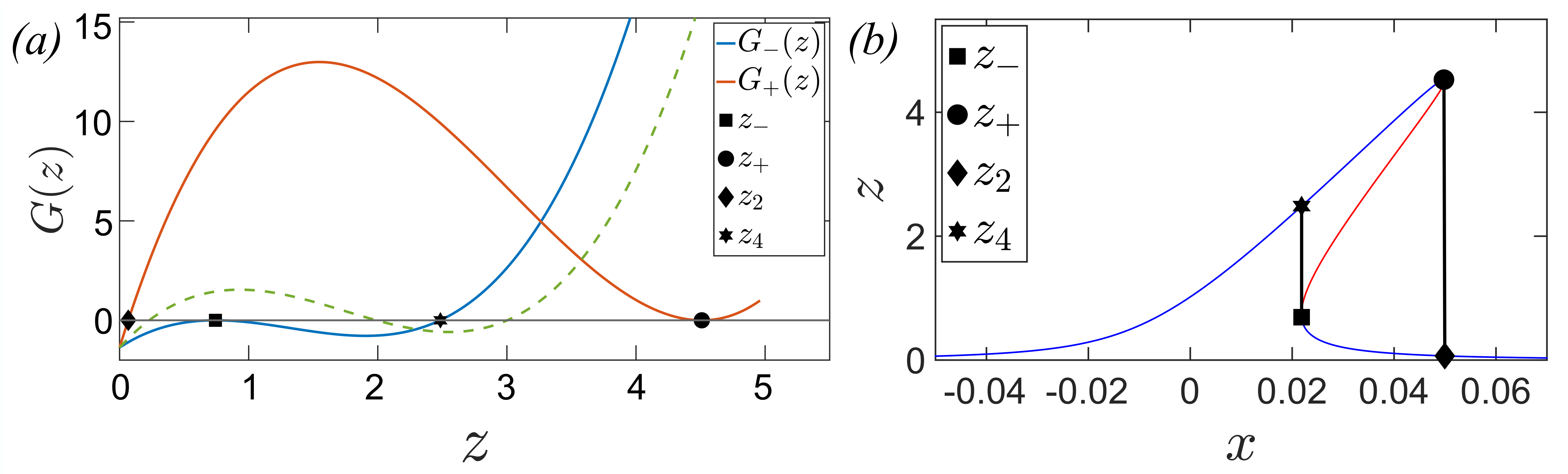}
\caption{\label{Fig:GofZ}
$(a)$ Function $G(z)$ for different values of the MDOF $x\propto A$. 
$G_{\pm}(z)$ stand for $G(z, x_{\pm})$. The green dashed line represents $G(z)$ for some arbitrary value 
of $x_{-}\leqslant x\leqslant x_{+}$. 
$(b)$ Hysteresis loop $z \propto|a|^{2}$ vs. $x$. By moving the MDOF from the left to the right, 
$x$ reaches the critical value $x_{+}$ where $z$ drops from $z_{+}$ (circle) down to $z_{2}$ (diamond)
and the MDOF continues moving to the right. On its way back, at critical value $x_{-}$, $z$ jumps from $z_{-}$ (square) to $z_{4}$ (star)
and the mirror continues its motion to the left and repeats. The red solid line corresponds to the third root of \cref{Eq:GofZ}, 
which is physically inaccessible.}
\end{figure}

In addition to a pathological case of no real roots in the domain of $z \geqslant 0$, there might be one or three real zeros
for the cubic equation \cref{Eq:GofZ}, as shown in \cref{Fig:GofZ}. The latter situation might occur only when $G(z)$ has 
one maximum (on the left) and one minimum (on the right), as shown by the green dashed line in \cref{Fig:GofZ}. The condition $G^{\prime}(z) = 0$ leads to a quadratic equation with solutions
\begin{equation}\label{Eq:Zpm}
z_{\pm} = \frac{2 A}{3} \pm\frac{1}{3}\sqrt{A^{2} - 3B}.
\end{equation}
Now, we notice that in \cref{Eq:GofZ} all the parameters are fixed by the setup parameters except for the variables $z$ and $x$,
which change their value during the system evolution in such a way that we may consider $z$ as a function of $x$ or vice versa.
Indeed, the photonic energy is a function of the mirror position, while the mechanical driving force and, therefore, the mirror position
is a function of photonic energy. In \cref{Eq:GofZ}, only the parameter $A$ depends on $x$. 
Here, we introduce $A_{\pm}$ as the values of $A$ that are solutions of the equations
\begin{equation}\label{Eq:Arange}
G[z_{\pm}(A_{\pm})] = 0.
\end{equation}
Explicitly,
\begin{equation}\label{Eq:Apm}
\left(\frac{2 A_{\pm}}{3} \pm\frac{1}{3}\sqrt{A_{\pm}^{2} - 3B}\right)
\left[\left(-\frac{A_{\pm}}{3} \pm \frac{1}{3}\sqrt{A_{\pm}^{2} - 3B}\right)^{2} + B\right] = J.
\end{equation}
Thus, $x_{\pm}\propto A_{\pm}$ are the values of the mirror position for which $G[z(x)]$ crosses the horizontal axes only twice, i.e., 
two of the three roots are degenerate (see \cref{Fig:GofZ}). Whenever $x < x_{-}$ or $x > x_{+}$, only one solution of \cref{Eq:GofZ} is possible.  
Therefore, $x_\pm$ indicate whether bistability is possible and they define the boundaries of the hysteresis loop.

\begin{figure}[ht]
\center\includegraphics[scale = .17]{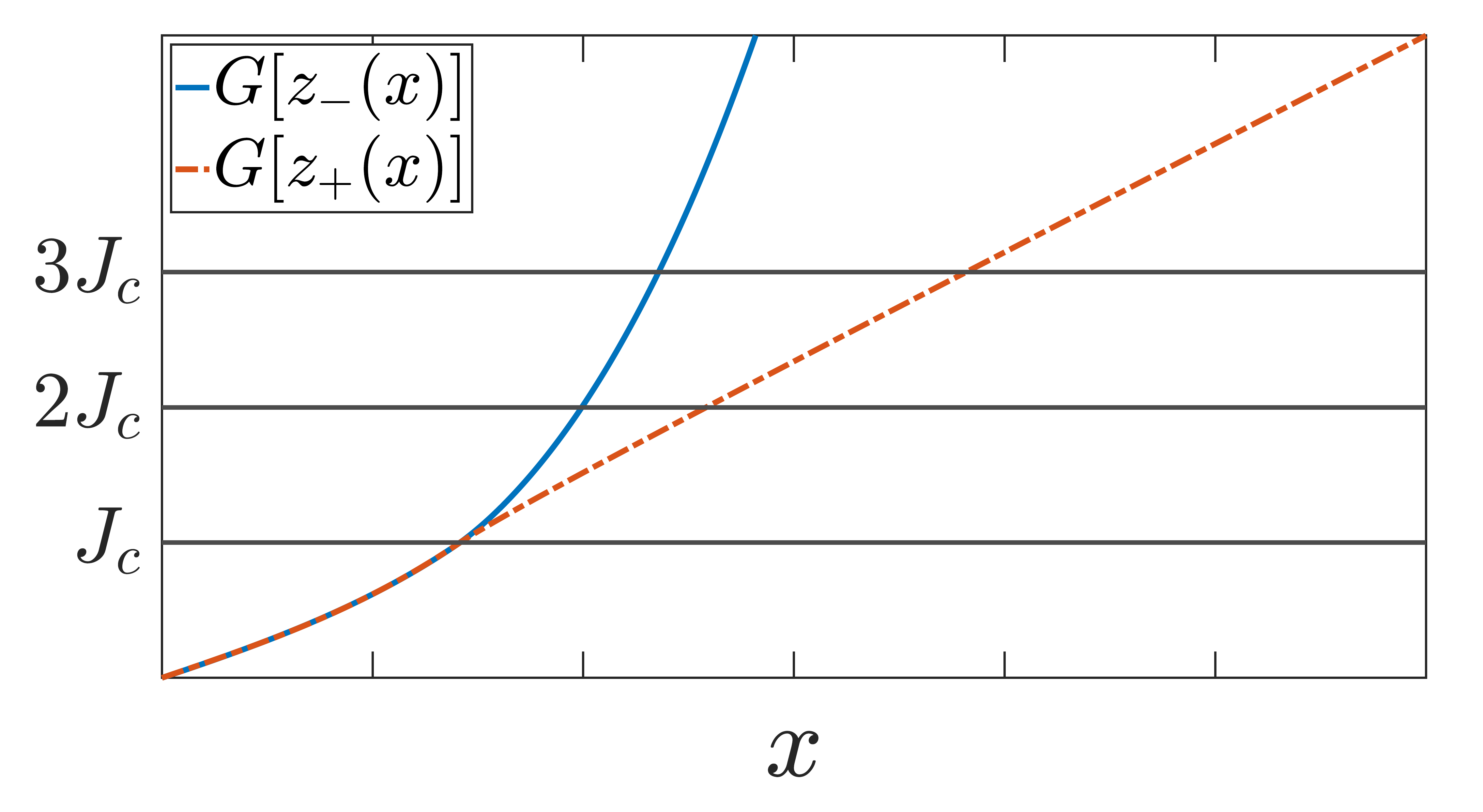}
\caption{\label{Fig:GplGmn}$G[z_{\pm}]$  vs $x$. At the critical value $J_{c}$,  \cref{Eq:Jc},
 two solutions of \cref{Eq:Arange} collapse into one point and no real solution below $J_{c}$ is possible 
(there is an imaginary component below $J_{c}$ not shown in the plot). By increasing $J$, we increase the width of the loop 
in \cref{Fig:GofZ}$(b)$.}
\end{figure}

By solving \cref{Eq:Apm},  we may deduce if there is a hysteresis loop for the given set of parameters. 
In such a case, the width of the hysteresis loop, which is proportional to the work of the motor after one cycle,
is entirely controlled by the parameter $J$ in \cref{Eq:GofZ}. 
There, we can identify a critical value of $z_{c} = z_{+} = z_{-}$ that causes the collapse of the bi-stability region, which
is defined from \cref{Eq:Zpm} as
\[
z_{c} = \frac{2A}{3}, \quad A^{2} = 3B.
\]
This leads to a critical value of $J$ (see \cref{Fig:GplGmn})
\begin{equation}\label{Eq:Jc}
J_{c} = \left[\frac{4 B}{3}\right]^{3/2}
\end{equation} 
above which bistability is possible. In turn, this implies that there is a critical value for the source power $|s_{p}^{cr}|^{2} $
to be overcome in order to have bistability. Using \cref{Eq:Params}, this condition can be formulated as
\begin{equation}\label{Eq:ScSingle}
|s_{p}^{}|^{2} > |s_{p}^{cr}|^{2} = \left(\frac{4}{3}\right)^{3/2} \frac{\gamma^{3}}{2\chi\gamma_{e}},
\end{equation}
for the single-cavity motor, while for the double-cavity motor one should use \cref{Eq:DCParams} to get 
\begin{equation}\label{Eq:Sc}
|s_{p}^{}|^{2} > |s_{p}^{cr}|^{2} =\left(\frac{4}{3}\right)^{3/2} \frac{2\bar{\gamma}^{6}}{\gamma_{e}^{}\gamma_{b}^{}\chi(\gamma_{a} -
\gamma_{b})^{2}}.
\end{equation}

\section{Mechanical Criterion\label{Sec:MechDoF}}

From the solution of \cref{Eq:SetUpRevisit}, we can predict whether work production is possible and also compute its value,
assuming that the whole bi-stability region is accessed by the MDOF. However, this assumption is not always satisfied. 
Indeed, by considering only the photonic equations of motion, we cannot obtain information whether $x$ will reach this 
region dynamically or not. Instead, we have to consider also \cref{Eq:3}. In this section, we discuss a criterion that
guarantees that the steady state motion of the MDOF will explore the whole bistability region, hence producing work.

The MDOF reaches its steady state when the dissipated energy $w_{\rm fr}$ is balanced by the work of the photonic force $w_{\rm ph}$
in each period of its motion.
Assuming resonant driving, the work done per period by the friction force results
\begin{equation}\label{Eq:Wfriction}
w_{\rm fr} = \int\limits_{0}^{2\pi/\Omega}2\Gamma \dot{x}^{2} dt =
 2\Gamma\Omega^{2} x_{0}^{2}\int\limits_{0}^{2\pi/\Omega}\cos^{2}\Omega t dt= 2\pi\Gamma\Omega x_{0}^{2}.
\end{equation}    
Here, we have assumed
\begin{equation}\label{Eq:Xoft}
x(t) = x_{\rm eq} + x_{0}\sin\Omega t,
\end{equation}
where $x_{\rm eq}\ll x_0$ is the equilibrium position of the MDOF.
Then, the amplitude $x_0$ is given by 
\begin{equation}\label{Eq:MechCrit}
x_{0} = \sqrt{\frac{w_{\rm ph}}{2\pi\Gamma\Omega}}, \quad w_{\rm ph} = \alpha\oint |a(x)|^{2}dx.
\end{equation}
If $x_{0} > \max\big\{|x_{-}|, |x_{+}|\big\}$, the MDOF covers the whole bistability region and work production 
can sustain a self-oscillation of the MDOF.

Notice that we have neglected the displacement of the equilibrium position of the MDOF since $x_{\rm eq}\ll x_{0}$. 
The precise criterion would be $x_{0} > \max\big\{|x_{-}|, |x_{+}|\big\} -  x_{\rm eq}$, however, we have numerically tested that 
neglecting $x_{\rm eq}$ does not affect the results.

\begin{figure}
\centering
\subfigure[ ]
{\label{Fig:ZvsXGb12Ga02EPx0}\includegraphics[scale = 0.25]{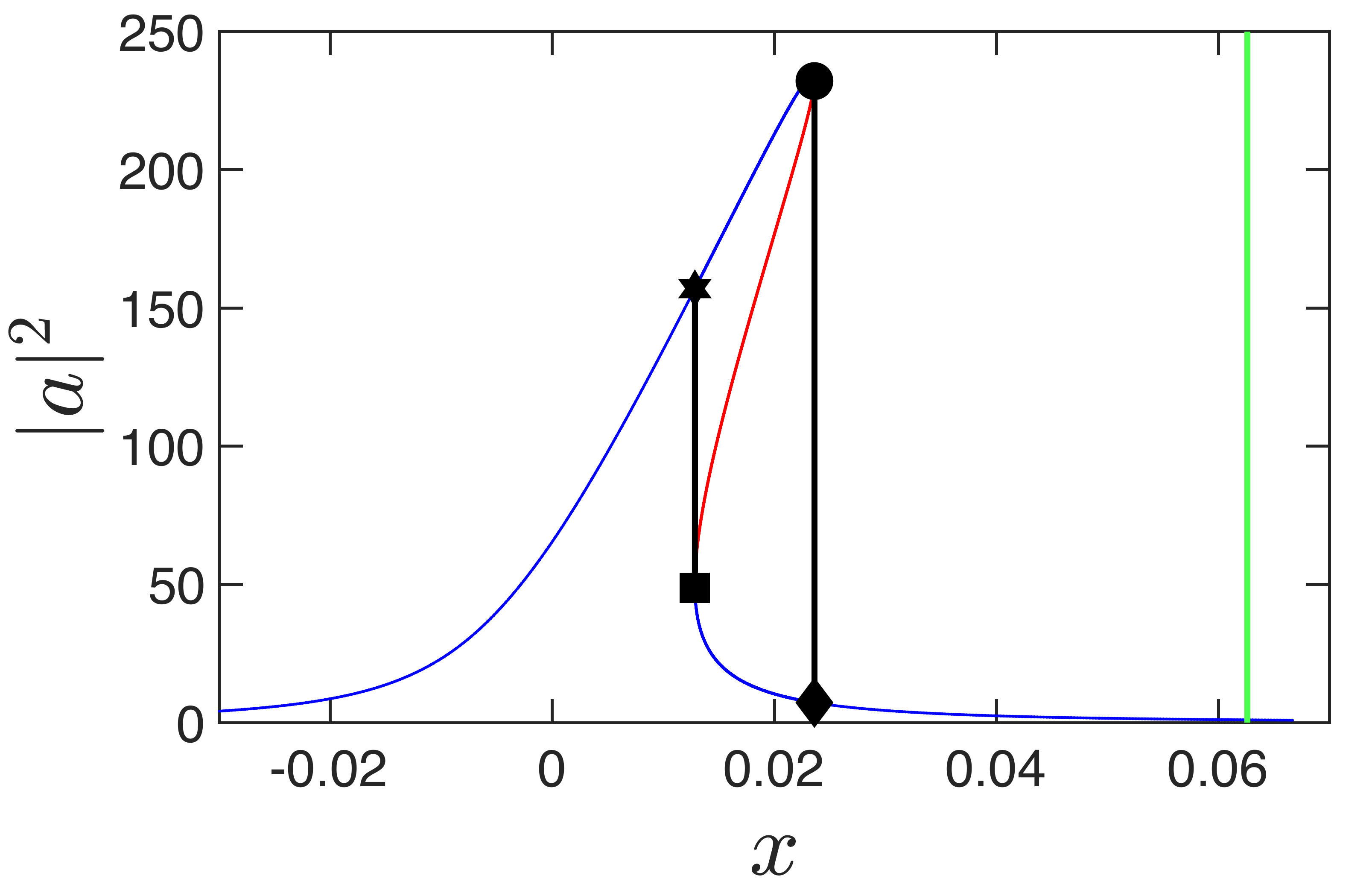}}
\subfigure[ ]{\label{Fig:ZvsXGb12Ga02OffEPx0}\includegraphics[scale = 0.25]{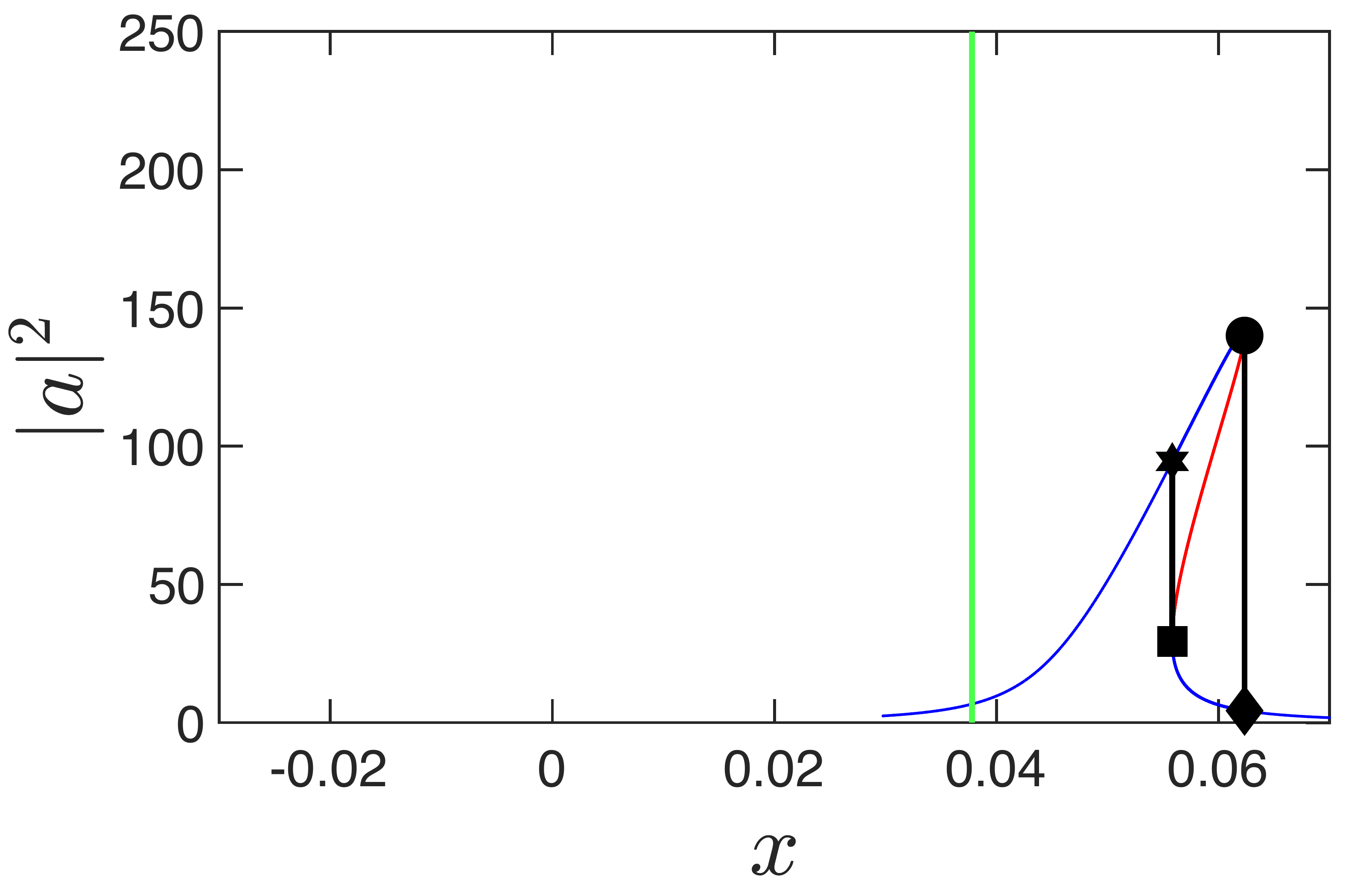}}
\caption{\label{Fig:ZvsXGb12Ga02theor} $|a|^{2}$ vs $x$ for $(a)$ $\varkappa =\sqrt{\gamma_{a}\gamma_{b}}$, $\delta\omega = 0$
and $(b)$ $\varkappa =2\sqrt{\gamma_{a}\gamma_{b}}$, $\delta\omega = -0.05\, \omega_{0}$. The rest of the parameters are the same,
$\omega_{0} = 1$, $|s_{\rm e}| = 0.15$, $\gamma_{a} = 2.0\cdot 10^{-3}$, $\gamma_{b} = 12\cdot 10^{-3}$, $\chi = 1.0\cdot 10^{-4}$, 
$\alpha = 1.25\cdot10^{-17}$. The green vertical line indicates the maximum amplitude of the MDOF, $x_0$.}
\end{figure}

As an example, in \cref{Fig:ZvsXGb12Ga02theor}, we consider two scenarios where we vary the coupling and emitter detuning,
given that the rest of the parameters are the same,
such that in both cases \cref{Eq:GofZ} predicts non-zero area of the hysteresis loop. 
Here, the MDOF covers the whole bistability region in case $(a)$ only, while for the case
in $(b)$, $x$ is not able to reach the value required to explore the two branches of the loop. In consequence, no work is produced after 
each cycle and finally the MDOF relaxes to the equilibrium position regardless of the initial conditions. 
We have verified these statements via dynamical simulations.

\section{\label{Sec:Isopower}Invariance of the work per cycle along the iso-power line.}

As discussed in the main text, the double-cavity setup has a nontrivial dependence of the mechanical power production 
with the laser detuning, $\delta\omega = \omega - \omega_{0}$, and the coupling, $\varkappa$. In the present section, 
we show that Eq. (9) of the main text (iso-power line), 
\begin{equation}\label{SUPEq:IsoPower}
\delta\omega^{2} = \frac{\gamma_{b}^{2}}{\varkappa_{0}^{2}}\left(\varkappa^{2} - \varkappa_{0}^{2}\right), 
\quad \varkappa \geqslant \varkappa_{0},
\end{equation}
indeed represents a line in the parameter space along which the power production is constant for any $\varkappa_{0} > 0$. 
 
In order to show this, we use again the fact that that the work per cycle is proportional to the area under the loop associated to 
the bistability region, e.g. in \cref{Fig:ZvsXGb12Ga02theor}. 
In principle, this hysteresis loop might be controlled by the parameters $A$, $B$, and $J$ in \cref{Eq:GofZ}, 
however, not all these parameters affect the area of the loop, and at the same time, some of these parameters
may remain unchanged under variations of the system parameters.
Indeed, one can verify that $B$ and $J$ are invariant under any values of $(\varkappa,\delta \omega)$ satisfying 
\cref{SUPEq:IsoPower}. To show this, we consider \cref{Eq:aux,Eq:GofZ,Eq:DCParams} which result in 
 \begin{equation*}
 \begin{split}
& \Gamma_{b}^{2} = \gamma_{b}^{2} + \frac{\gamma_{b}^{2}}{\varkappa_{0}^{2}}\left(\varkappa^{2} - \varkappa_{0}^{2}\right) = 
 \gamma_{b}^{2}\frac{\varkappa^{2}}{\varkappa_{0}^{2}}\\
 &\eta^{2} = \left(\varkappa_{0}^{4} + \gamma_{a}^{2}\gamma_{b}^{2}\right)\frac{\varkappa^{2}}{\varkappa_{0}^{2}}
 \end{split}
 \end{equation*}
 and therefore 
  \begin{equation*}
 \begin{split}
&B = \frac{\varkappa_{0}^{4} + \gamma_{a}^{2}\gamma_{b}^{2}}{\bar{\gamma}^{2}\gamma_{b}^{2}}\\
&J = \frac{\varkappa_{0}^{2}\chi|s_{e}|^{2}}{\bar{\gamma}^{3}\gamma_{b}^{2}}.
 \end{split}
 \end{equation*}
Here, $B$ and $J$ are invariant under variations of $(\varkappa,\delta \omega)$ satisfying the condition imposed by \cref{SUPEq:IsoPower}. 
Finally, the parameter $A$ may shift the the boundaries of the hysteresis loop $x_{\pm}$, but the bistability region 
$A_{-}\leqslant A\leqslant A_{+}$, which is determined entirely by values of $B$ and $J$, remains unchanged as well as its area.
This invariance is translated to the mechanical work, unless the required values of $x$ become unsupported by the MDOF due to the
mechanical criterion, see \cref{Sec:MechDoF}.
 
It could be shown that, for a given value of $\gamma_{b}$, the curves from the parametric family of \cref{SUPEq:IsoPower} do not cross.
This result suggests a simple method to determine the maximum power production on the manifold $(\delta\omega, \varkappa)$,
assuming the same set of the other parameters. 
Indeed, it is enough to find a maximum value of the mechanical power production along the line $\delta\omega = 0$ and 
the corresponding value of coupling $\varkappa_{\ast}$. This power value is the maximum possible and it is preserved constant 
for any pair of detuning and coupling values bound by  \cref{SUPEq:IsoPower}  with $\varkappa_{0} = \varkappa_{\ast}$.

\section{\label{Sec:Noise}Noise quantification}

In this section, we provide details about the noise strength quantification.
In our time-domain simulations, we consider white noise as a stochastic component in addition to 
the monochromatic driving in \cref{Eq:SC1,Eq:1}. To quantify the noise strength as compared to the signal, we consider
a single mode $a$ whose dynamics is given by a Langevin equation,
\begin{equation}\label{Eq:LinearNoise}
\frac{d a}{dt} = i\omega_{r}a - \gamma a  + 
i s_{p}\sqrt{2\gamma_{e}} e^{i\omega t} + i\sqrt{2\gamma_{n}\theta}\xi(t).
\end{equation} 
Here, $\omega_r$ is the resonant frequency, $\gamma$ represents the total loss, $|s_p|^2$ and $\omega$ are the power 
and the frequency of the driving source, $\theta$ is the effective temperature (in energy units),
and $\xi(t)$ is delta-correlated Gaussian stochastic process:
\[
\langle\xi^{\ast}(t)\xi(t^{\prime})\rangle = \delta(t - t^{\prime}), \quad \langle\xi(t)\rangle = 0.
\]
For simplicity, we assume $\gamma_{n} = \gamma_{e} = \gamma$. 
For further analysis, it is convenient to rewrite \cref{Eq:LinearNoise} in dimensionless form:
\begin{equation}\label{Eq:NoiseDimensionless}
\frac{d a_{u}}{d\tau} = i a_{u} - \gamma_{u} a_{u}  + 
ie^{i\varepsilon \tau} + i\lambda\xi(\tau),
\end{equation} 
where
\begin{equation}
\begin{split}
&\varepsilon = \frac{\omega}{\omega_{r}}, \quad \gamma_{u} = \frac{\gamma}{\omega_{r}},\\
&a_{u} = a\frac{\omega_{r}}{\sqrt{2\gamma_{e}|s_{p}|^{2}}}, \quad \lambda = 
\sqrt{\frac{\gamma_{n}}{\gamma_{e}}}\sqrt{\frac{\omega_{r} \theta}{|s_{p}|^{2}} },\\ 
&\langle\xi^{\ast}(\tau)\xi(\tau^{\prime})\rangle = \delta(\tau - \tau^{\prime}), \quad \langle\xi(\tau)\rangle = 0.
\end{split}
\end{equation}
%
\cref{Eq:NoiseDimensionless} could be formally integrated as 
\begin{equation}\label{Eq:au}
a_{u}(\tau) = ie^{(i - \gamma_{u})\tau}\int\limits_{0}^{\tau}ds\Big\{e^{[\gamma_{u} + i(\varepsilon - 1)]s} + \lambda e^{(\gamma_{u} - i)s}\xi(s)\Big\}
\end{equation}
where we assume $a_{u}(0) = 0$. The average energy stored in the mode is proportional to the dimensionless value of
$\mathcal{E} = \langle\, |a_{u}(\tau)|^{2}\rangle$  and could be found from \cref{Eq:au} and $\xi$'s correlations properties as
\begin{equation}\label{Eq:absA}
\langle\, |a_{u}(\tau)|^{2}\rangle \xrightarrow{\tau\gg \gamma_{u}^{-1}}  
\frac{1}{(\varepsilon - 1)^{2} + \gamma_{u}^{2}} + \frac{\lambda^{2}}{2\gamma_{u}} .
\end{equation}
The two terms in the latter expression correspond to regular, $\mathcal{E}_{r}$, and stochastic, $\mathcal{E}_{s}$, components respectively. 
In the resonant absorption scenario $\varepsilon=1$, the regular component is dominant. Indeed, for typical values
of $\lambda \sim 1$ and $\gamma_{u} = 0.01$
the contribution ratio is $\mathcal{E}_{r}/\mathcal{E}_{s}\approx \lambda^{2}/2\gamma_{u}\sim 50$. 

Here, we quantify the noise strength via the fluctuations of the modal energy, given by the variance 
\begin{equation}\label{Eq:Variance}
\sigma_{a}^{2} = \mathrm{Var} (|a_{u}(\tau)|^{2}) = \langle|a_{u}(\tau)|^{4}\rangle - \langle|a_{u}(\tau)|^{2}\rangle^{2} = \\
\lambda^{2}\gamma_{u}^{-3} + \mathcal{O}(\gamma_{u}^{-2}).
\end{equation}
Here, the four-point correlation is evaluated using Isserlis' theorem and only the main asymptotics of $\gamma_{u}^{-1}$ in 
the final expression preserved. Finally, we quantify the noise strength by the expression
\begin{equation}\label{Eq:NS}
\mathrm{NS} = \frac{\sigma_{a}}{\mathcal{E}}\approx \frac{\lambda\gamma_{u}^{-3/2}}{\gamma_{u}^{-2}} = \lambda\gamma_{u}^{1/2}.
\end{equation}
Therefore, typical parameters used in the main text are $\gamma/\omega_{0} = 0.012$ ($\gamma_{b}/\omega_{0}$ 
in the double-cavity motor) and noise amplitude $\lambda = 1$ that correspond to a noise strength $\mathrm{NS}\approx 10\%$.
This amount of noise represents huge fluctuations as compared with the ones associated to typical quantum noise values in lasers. 

\section{Comparison with existing self-oscillations in optomechanical systems.}

In this section, we compare our mechanism with classical nonlinear
dynamics in the context of optomechanics \cite{SUPaspelmeyer2014}. In this
case, the coupled equations of motion between the photonic and mechanical
degrees of freedom, as presented in Ref. \onlinecite{SUPMarquardt2006}, are
\begin{eqnarray}
\dot{a}_{m}\left(t\right) & = & \left(ix_{m}-\frac{1}{2}\right)a_{m}\left(t\right)+\frac{1}{2}\label{eq:dot_alpha}\\
\ddot{x}_{m} & = & \mathcal{P}\left|a_{m}\right|^{2}-\omega_{m}^{2}\left(x_{m}-x_{m}^{0}\right)-\Gamma_{m}\dot{x}_{m}\label{eq:ddot_x}
\end{eqnarray}
Here, $x_{m}\left(t\right)$ and $a_{m}\left(t\right)$ describe
the MDOF and the PDOF respectively. The latter is given by 
$a_{m}\left(t\right)=\frac{\widetilde{a}_{m}e^{i\omega_{L}t_{m}}}{\sqrt{n_{\mathrm{max}}}}$,
where $t_{m}$ is the time measured in units of the ring-down time
of the cavity $\gamma_{m}^{-1}$, $\omega_{L}$ is the laser frequency,
$\widetilde{a}_{m}$ is the coherent light amplitude, and $n_{\mathrm{max}}$
is the maximum photon number ($n_{\mathrm{max}}=4P_{in}/(\gamma\hbar\omega_{L}),$where
$P_{in}$ is the input power). The parameters of the MDOF $\mathcal{P}$,
$\omega_{m}$, $x_{m}^{0}$, and $\Gamma_{m}$ provide, respectively,
the coupling with the PDOF, the unperturbed resonant frequency,
the unperturbed equilibrium position and the mechanical damping
rate.

The dynamics of $a\left(t\right)$ resembles that of a driven damped
oscillator with a natural frequency that is swept through resonance
non-adiabatically. Typically, the effects of radiation per cycle
are weak, such that $x_{m}\left(t\right)$ moves approximately
with sinusoidal oscillations, $x_{m}=\overline{x}_{m}+A_{m}\cos\left(\omega_{m}t_{m}\right)$,
where $A_{m}$ is the amplitude of the oscillation.

In Ref. \onlinecite{SUPMarquardt2006}, the authors provide and
analytical expression for the output power $P_{\mathrm{out}}^{(m)}$
produced by the mechanical force ($P_{\mathrm{out}}^{(m)}=\mathcal{P}\left\langle \left|a_{m}\right|^{2}\dot{x}_{m}\right\rangle _{t_{m}}$),
which reads 
\begin{eqnarray*}
P_{\mathrm{out}}^{(m)} & = & \mathcal{P}A_{m}\omega_{m}\mathrm{Im}\left[\sum_{n}a_{m}^{*}\left(n,\omega_{m}\right)a_{m}\left(n+1,\omega_{m}\right)\right].
\end{eqnarray*}
Here, the supra-index $(m)$ indicates that the output power is in
units consistent with Eqs. \ref{eq:dot_alpha} and \ref{eq:ddot_x}
and $a_{m}\left(n,\omega_{m}\right)=\frac{1}{2}\frac{J_{n}\left(-\frac{A_{m}}{\omega_{m}}\right)}{in\omega_{m}+\frac{1}{2}-i\overline{x}}$
where $J_{n}$ is the Bessel function of the first kind. The above
equation can be recast as
\begin{eqnarray*}
P_{\mathrm{out}}^{(m)} & = & \sum_{n}^{\infty}\frac{\left(-\mathcal{P}\frac{A_{m}}{8}\omega_{m}^{2}\right)J_{n}\left(-\frac{A_{m}}{\omega_{m}}\right)J_{n+1}\left(-\frac{A_{m}}{\omega_{m}}\right)}{\left(n\left(n+1\right)\omega_{m}^{2}-\left(2n+1\right)\overline{x}\omega_{m}+\overline{x}^{2}+\frac{1}{4}\right)^{2}+\left(\frac{\omega_{m}}{2}\right)^{2}}.
\end{eqnarray*}
By taking Eqs. \ref{eq:dot_alpha} and \ref{eq:ddot_x} and performing
the following replacements: $a=\widetilde{a}_{m}\sqrt{\hbar\omega_{L}}$,
$t_{m}=t\gamma$, $\overline{x}_{m}=x_{\mathrm{m}}^{0}+\delta x_{m}^{0}$,
$X=\left[\frac{\left(-\delta x_{\mathrm{m}}^{0}\right)}{x_{\mathrm{m}}^{0}}+\frac{\left(-A\right)}{x_{\mathrm{m}}^{0}}\cos\left(\omega_{m}\gamma_{m}t\right)\right]$,
$\omega_{0}=\left(x_{\mathrm{m}}^{0}\gamma_{m}\right)$, $\gamma=\frac{\gamma_{m}}{2}$,
$P_{in}=\left|s_{p}\right|^{2}$, $\omega=\omega_{L}\gamma_{m}$,
$\Gamma=\frac{\gamma_{m}\Gamma_{m}}{2}$, $\Omega=\gamma_{m}\omega_{m}$
and $\alpha=\left(-\frac{\gamma_{m}^{3}\mathcal{P}}{4x_{\mathrm{m}}^{0}P_{in}}\right)$
we recover exactly our Eqs. 3(a,b) of the main text (but without the nonlinear photonic term).

The output power, in units consistent with our equations ($P_{\mathrm{out}}=\frac{1}{\tau}\intop_{0}^{\tau}\alpha\left|a\right|{}^{2}\dot{X}dt$)
and written in terms of our parameters, is
\begin{eqnarray*}
P_{\mathrm{out}} & = & \sum_{n}^{\infty}\frac{\left(-\frac{\alpha\left|s_{p}\right|^{2}}{2x_{0}}\right)\left(\frac{\Omega}{\omega_{0}}\right)^{2}J_{n}\left(-\frac{x_{0}}{\left(\Omega/\omega_{0}\right)}\right)J_{n+1}\left(-\frac{x_{0}}{\left(\Omega/\omega_{0}\right)}\right)}{\left(n\left(n+1\right)\frac{\omega_{0}}{2\gamma x_{0}}\left(\frac{\Omega}{\omega_{0}}\right)^{2}-\left(2n+1\right)\frac{\omega_{0}}{2\gamma x_{0}}\left(\frac{\Omega}{\omega_{0}}\right)+\frac{\omega_{0}}{2\gamma x_{0}}+\frac{2\gamma}{4x_{0}\omega_{0}}\right)^{2}+\frac{1}{4x_{0}^{2}}\left(\frac{\Omega}{\omega_{0}}\right)^{2}}.
\end{eqnarray*}
Here, we emphasize that in Ref. \onlinecite{SUPMarquardt2006}, the authors were using a completely different approach and they were 
treating the amplitude of the MDOF's oscillation $x_0$ as a parameter while, in our case, $x_0$ corresponds to a single value
that results from energy conservation.

\begin{figure}
\begin{centering}
\includegraphics[width=3.5in]{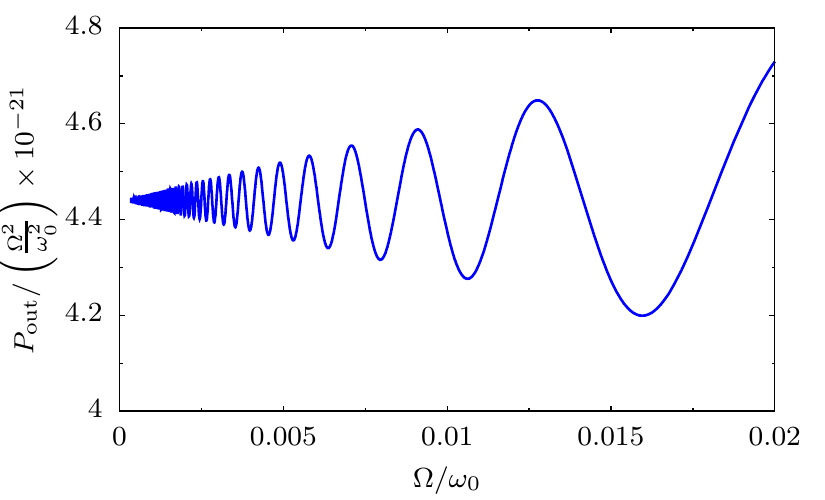}
\par\end{centering}
\caption{Output power divided by $\left(\Omega^{2}/\omega_{0}^{2}\right)$
as function of the frequency of the MDOF $\Omega/\omega_{0}$. For
this plot we used the parameters indicated in Ref. \onlinecite{SUPkippenberg2005}
and $x_{0}=0.1$ which is of the order of the amplitudes
shown in panels ($c$) and ($d$) of Fig. 1 of the main text. \label{fig:limit_Omega}}
\end{figure}

Although not so obvious from the equations, numerical evaluations confirm
that the output power goes as $\Omega^{2}$ in the limit of small
$\Omega$, see Fig. \ref{fig:limit_Omega}. This has at least two
main consequences. First, it is clear then that, in the adiabatic
limit, only our mechanism can contribute significantly to the output power, since
it is linear in $\Omega$. Second, without a nonlinear photonic term,
self-oscillation with large frequency separation between the MDOF
and PDOF is only possible for high quality factors, otherwise, the
expected oscillation amplitude will be extremely small. As can be
seen in the specialized literature, this is indeed the case. For
example, in Ref. \onlinecite{SUPCarmon05} the authors observed
self-oscillations but using cavities with a value of $\gamma$ five
orders of magnitude smaller than the ones used in the present manuscript,
while in Ref. \onlinecite{SUPMarquardt2006} they used a value
of $\gamma$ three orders of magnitude smaller.


\begin{thebibliography}{1}

\bibitem{CKOF16}B. S. L. Collins, J. C. M. Kistemaker, E. Otten, B. L. Feringa, {\it A chemically powered 
unidirectional rotary molecular motor based on a palladium redox cycle.}, Nat. Chem {\bf 8}, 860 (2016).

\bibitem{WSCGLL16}M. R. Wilson, J. Sol\'a, A. Carlone, S. M. Goldup, N. Lebrasseur, D. A. Leigh, {\it An 
autonomous chemically fuelled small-molecule motor}, Nature (London) {\bf 534}, 235 (2016).

\bibitem{KZDHF99}N. Koumura, R. W. Zijlstra, R. A. van Delden, N. Harada, and B. L. Feringa, {\it Light-
driven monodirectional molecular rotor}, Nature (London) {\bf 401}, 152 (1999).


\bibitem{TMJBKMKS11} H. L. Tierney, C. J. Murphy, A. D. Jewell, A. E. Baber, E. V. Iski, H. Y. Khodaverdian, 
A. F. McGuire, N. Klebanov, and E. C. H. Sykes, {\it Experimental demonstration of a single-molecule electric 
motor}, Nat. Nanotechnol. {\bf 6}, 625 (2011).

\bibitem{KRPMKHEF11}T. Kudernac, N. Ruangsupapichat, M. Parschau, B. Maci\'a, N. Katsonis, S. R. Harutyunyan, K.-H. 
Ernst, and B. L. Feringa, {\it Electrically driven directional motion of a four-wheeled molecule on a metal surface},
Nature (London) {\bf 479}, 208 (2011).


\bibitem{LCGJC20} H. H. Lin, A. Croy, R. Gutierrez, C. Joachim, G. Cuniberti, 
{\it Mechanical transmission of rotational motion between molecular-scale gears},
Phys. Rev. Applied {\bf 13}, 034024 (2020).


\bibitem{PG13} D. Palima, J. Gl\"uckstad, {\it Gearing up for optical microrobotics: micromanipulation and actuation 
of synthetic microstructures by optical forces}, Laser Photon. Rev. {\bf 7}, 478 (2013).

\bibitem{CSGWSSBGPHM10}D. M. Carberry, S. H. Simpson, J. A. Grieve, Y. Wang, H. Sch\"afer, M. Steinhart, R. Bowman,
 G. M. Gibson, M. J. Padgett, S. Hanna, M. J. Miles, {\it Calibration of optically trapped nanotools}, Nanotechnology {\bf 21}, 
175501 (2010).

\bibitem{WNMMMRB12} T. Wu, T. A. Nieminen, S. Mohanty, J. Miotke, R. L. Meyer, H. Rubinsztein-Dunlop, M. W. 
Berns, {\it A photon-driven micromotor can direct nerve fibre growth}, Nature Photon. {\bf 6}, 62 (2012).

\bibitem{bustos2013}
R. Bustos-Mar\'un, G. Refael, and F. von Oppen, \textit{Adiabatic quantum motors}, Phys. Rev. Lett. \textbf{111}, 060802 (2013).


\bibitem{CCBE16} A. Celestino, A. Croy, M. W. Beims, A. Eisfeld, 
{\it Rotational directionality via symmetry-breaking in an electrostatic motor}, New J. Phys. {\bf 18}, 063001 (2016).


\bibitem{GK14}  D. Gelbwaser-Klimovsky and G. Kurizki, {\it Work extraction from heat-powered quantized optomechanical setups}, Sci. Rep. {\bf 5}, 07809 (2015).

\bibitem{RKSCPCP18} A. Ronzani, B. Karimi, J. Senior, Y.-C. Chang, J. Peltonen, C.D. Chen, and J. P. Pekola. {\it Tunable photonic heat transport in a quantum heat valve}, Nature Phys. {\bf 14}, 991 (2018).

\bibitem{ZBM14} K. Zhang, F. Bariani, and P. Meystre. {\it Quantum Optomechanical Heat Engine}. Phys. Rev. Lett. {\bf 112}, 150602 (2014).

\bibitem{KBH21} S. Khandelwal , N. Brunner, and G. Haack, {\it Signatures of Liouvillian Exceptional Points in a Quantum Thermal Machine},
Phys. Rev. Quantum {\bf 2}, 040346 (2021).

\bibitem{KSMLR14} S. Kheifets, A. Simha, K. Melin, T. Li, M. G. Raizen, {\it Observation of Brownian motion in 
liquids at short times: instantaneous velocity and memory loss}, Science {\bf 343}, 1493 (2014).

\bibitem{GFMLCD16}M. Serra-Garcia, A. Foehr, M. Moleron, J. Lydon, C. Chong, C. Daraio, {\it Mechanical Autonomous 
Stochastic Heat Engine}, Phys. Rev. Lett. {\bf 117}, 010602 (2016).



\bibitem{RS08} D. C. Ralph and M. D. Stiles, {\it Spin Transfer Torques}, J. Magn. Magn. Mater. {\bf 320}, 1190 (2008).

\bibitem{AvO15} L. Arrachea, F. von Oppen, {\it Nanomagnet coupled to quantum spin Hall edge: An adiabatic quantum motor},
Physica E: Low-dimensional Systems and Nanostructures {\bf 74}, 596-602 (2015).

\bibitem{NBLACBS06} A. Naik, O. Buu, M. D. LaHaye, A. D. Armour, A. A. Clerk, M. P. Blencowe, K. C. Schwab,
{\it Cooling a nanomechanical resonator with quantum back-action}, Nature (London) {\bf 443}, 193-196 (2006).

\bibitem{STPBXPWBR10} J. Stettenheim, M. Thalakulam, F. Pan, M. Bal, Z. Ji, W. Xue, L. Pfeiffer, K. W. West, M. P. Blencowe, A. J. Rimberg,
{\it A macroscopic mechanical resonator driven by mesoscopic electrical back-action}, Nature (London) {\bf 466}, 86 (2010).


\bibitem{BMC10} S. D. Bennett, J. Maassen, and A. A. Clerk, 
{\it Scattering Approach to Backaction in Coherent Nanoelectromechanical Systems}, Phys. Rev. Lett. {\bf 105}, 217206 (2010).

 

\bibitem{DMT09} D. Dundas, E. J. McEniry, and T. N. Todorov, {\it Current-driven atomic waterwheels}, Nat. Nanotech. {\bf 4}, 99 (2009).

\bibitem{TDM10} T. N. Todorov, D. Dundas, E. J. McEniry, {\it Nonconservative generalized current-induced forces}, 
Phys. Rev. B {\bf 81}, 075416 (2010).

\bibitem{BKVEO11} N. Bode, S. Kusminskiy, R. Egger, F. von Oppen, {\it Scattering Theory of Current-Induced Forces in Mesoscopic Systems}, Phys. Rev. Lett.  {\bf 107}, 036804 (2011).

\bibitem{FBP15} L. J. Fern\'andez-Alc\'azar, R. A. Bustos-Mar\'un, H. M. Pastawski, {\it Decoherence in current induced forces: Application to adiabatic quantum motors}, Phys. Rev. B {\bf 92 } 075406 (2015).

\bibitem{brouwer1998}
P. W. Brouwer, \textit{Scattering approach to parametric pumping}, {Phys. Rev. B} \textbf{58}, R10135 (1998).


\bibitem{BTTOFA20}
B. Bhandari, P. T. Alonso, F. Taddei, F. von Oppen, R. Fazio, and L. Arrachea, \textit{Geometric properties of adiabatic quantum thermal 
machines}, Phys. Rev. B \textbf{102}, 155407 (2020).


\bibitem{BS20} K. Brandner, K. Saito, {\it Thermodynamic Geometry of Microscopic Heat Engines}, Phys. Rev. Lett.
{\bf 124}, 040602 (2020).


\bibitem{PPSW18} B. A. Placke, T. Pluecker, J. Splettstoesser, M. R. Wegewijs, 
{\it Attractive and driven interactions in quantum dots: Mechanisms for geometric pumping},
Phys. Rev. B {\bf 98}, 085307 (2018).

\bibitem{BC19} R. A. Bustos-Marun and H. L. Calvo, {\it Thermodynamics and steady state of quantum motors and 
pumps far from equilibrium}, Entropy {\bf 21}, 824 (2019).

\bibitem{SPTZ10} J. S. Seldenthuis, F. Prins, J. M. Thijssen, H. S. J. van der Zant, 
{\it An All-Electric Single-Molecule Motor}, ACS Nano {\bf 4}, 6681-6686 (2010).

\bibitem{KXGF14} K. Kim, X. Xu, J. Guo, D. L. Fan, {\it Ultrahigh-speed rotating nanoelectromechanical system devices assembled from nanoscale building blocks}, Nature Communications {\bf 5}, 3632 (2014).

\bibitem{FYHFCZ03} A. M. Fennimore, T. D. Yuzvinsky, Wei-Qiang Han, M. S. Fuhrer, J. Cumings, A. Zettl,
{\it Rotational actuators based on carbon nanotubes}, Nature {\bf 424}, 408-410 (2003).


\bibitem{FKK21} 
L. J. Fern\'andez-Alc\'azar, R. Kononchuk, T. Kottos, 
{\it Enhanced energy harvesting near exceptional points in systems with (pseudo-)PT-symmetry},
Communications Physics {\bf 4}, 79 (2021).



%


\bibitem{NEGRE2008} C. F. A. Negre, P. A. Gallay, C. G. Sánchez, {\it Model non-linear nano-electronic device},
Chem. Phys. Lett. {\bf 460}, 220 (2008)

\bibitem{DFT64} P. Hohenberg, W. Kohn, {\it Inhomogeneous electron gas}, Phys. Rev. {\bf 136}, B864 (1964). 

\bibitem{LHFL09} V. Loriot, E. Hertz, O. Faucher, and B. Lavorel, 
{\it Measurement of high order Kerr refractive index of major air components}, Opt. Express {\bf 17}, 13429 (2009).

\bibitem{film06} N. Moll, S. Jochim, S. Gulde,R. F. Mahrt,B. J. Offrein, {\it Organic nonlinear Kerr materials in Fabry-Perot 
cavities for all optical switching}, Photonic Crystal Materials and Devices IV {\bf 6128}, 202 (2006) 

\bibitem{aspelmeyer2014} M. Aspelmeyer, T. J. Kippenberg, F. Marquardt \textit{Cavity optomechanics}, Rev. Mod. Phys. \textbf{86}, 1391 
(2014).


\bibitem{SuppM} See the Supplementary Material for discussions about analytical solutions
of the Eqs. \ref{Eq:SingleCavity}, \ref{Eq:DoubleCavity}, \ref{MechCrit}, \ref{Eq:Isopower}, and the noise strength quantification.


\bibitem{G99}
E.~Gluskin, \textit{{A nonlinear resistor and nonlinear inductor using a
  nonlinear capacitor}}, J. Franklin Inst. \textbf{336}, 1035 (1999).

\bibitem{FKLK21} L. J. Fern\'andez-Alc\'azar, R. Kononchuk, H. Li, T. Kottos, 
{\it Extreme Non-Reciprocal Near-Field Thermal Radiation via Floquet Photonics},
Phys. Rev. Lett. {\bf 126}, 204101 (2021).




\bibitem{guha2020} B. Guha, P. E. Allain, A. Lemaitre, G. Leo, I. Favero, \textit{Force Sensing with an Optomechanical Self-Oscillator}, Phys. 
Rev. Applied \textbf{14}, 024079 (2020). 

\bibitem{zaitsev2011} S. Zaitsev, A. K. Pandey, O. Shtempluck, and E. Buks, \textit{Forced and self-excited oscillations of an optomechanical 
cavity} Phys. Rev. E \textbf{84}, 046605 (2011).

\bibitem{rokhsari2005} Rokhsari, H., T. J. Kippenberg, T. Carmon, and K. J. Vahala, \textit{Radiation-pressure-driven micro-mechanical 
oscillator }, Opt. Express \textbf{13}, 5293 (2005).


\bibitem{kippenberg2005}
T. J. Kippenberg, H. Rokhsari, T. Carmon, A. Scherer, and K. J. Vahala, \textit{Analysis of Radiation-Pressure Induced Mechanical Oscillation 
of an Optical Microcavity}, Phys. Rev. Lett. \textbf{95}, 033901 (2005).


\bibitem{Carmon05}
T. Carmon, H. Rokhsari, L. Yang, T. J. Kippenberg, and K. J. Vahala, 
{\textit Temporal Behavior of Radiation-Pressure-Induced Vibrations of an Optical Microcavity Phonon Mode},
Phys. Rev. Lett. {\bf 94}, 223902 (2005).

\bibitem{Marquardt2006} 
F. Marquardt, J. G. E. Harris, and S. M. Girvin, {\it Dynamical Multistability Induced by Radiation Pressure in High-Finesse Micromechanical Optical Cavities}, Phys. Rev. Lett. 96, 103901 (2006).

\bibitem{Hohenberg04}
H\"ohberger, C., and K. Karrai, {\it Self-oscillation of micromechanical resonators},
Proceedings of the 4th IEEE Conference on Nanotechnology (IEEE, New York) (2004).


\bibitem{jenkins2013}
A. Jenkins. \textit{Self-oscillation.} {Phys. Rep.} \textbf{525}, 167 (2013).

\bibitem{serragarcia2018} M. Serra-Garcia, M. Moler\'on, and C. Daraio.
\textit{Tunable synchronized down-conversion in magnetic lattices with defects},
Phil. Trans. Roy. Soc. A: Math. Phys. Eng. Sci. \textbf{376}, 2127 (2018).

\bibitem{NGS08} C. F. A. Negre, P. A. Gallay, C. G. S\'anchez, {\it Model non-linear nano-electronic device},
Chemical Physics Letters {\bf 460}, 220-224 (2008).

\bibitem{BALNO12} N. Bode, L. Arrachea, G. S. Lozano, T. S. Nunner, F. von Oppen, 
{\it Current-induced switching in transport through anisotropic magnetic molecules},
Phys. Rev. B {\bf 85}, 115440 (2012).



\bibitem{param1}
Parameters: $\omega_0 = 1.0$,  $\gamma_a = 2.0\cdot10^{-3}$,  $s_p = 2.0$ and $s_p=1.7$ (inset), $\Omega = 1.0\cdot 10^{-5}$, $\Gamma = 
\Omega/100$, $\alpha = (1.25/4)\cdot10^{-14}$. $(d)$ For the single-cavity, we use the same parameters and 
  $\gamma_e = \gamma =\bar{\gamma} \equiv (\gamma_a + \gamma_b)/2$.



\end{thebibliography}

\begin{thebibliography}{1}

\bibitem{SUPaspelmeyer2014} M. Aspelmeyer, T. J. Kippenberg, F. Marquardt \textit{Cavity optomechanics}, 
Rev. Mod. Phys. \textbf{86}, 1391 (2014).
  
\bibitem{SUPMarquardt2006} 
F. Marquardt, J. G. E. Harris, and S. M. Girvin, {\it Dynamical Multistability Induced by Radiation Pressure in High-Finesse Micromechanical Optical Cavities}, Phys. Rev. Lett. 96, 103901 (2006).

\bibitem{SUPCarmon05}
T. Carmon, H. Rokhsari, L. Yang, T. J. Kippenberg, and K. J. Vahala, 
{\textit Temporal Behavior of Radiation-Pressure-Induced Vibrations of an Optical Microcavity Phonon Mode},
Phys. Rev. Lett. {\bf 94}, 223902 (2005).

\bibitem{SUPkippenberg2005}
T. J. Kippenberg, H. Rokhsari, T. Carmon, A. Scherer, and K. J. Vahala, \textit{Analysis of Radiation-Pressure Induced Mechanical Oscillation 
of an Optical Microcavity}, Phys. Rev. Lett. \textbf{95}, 033901 (2005).



\end{thebibliography}

\end{document}